\documentclass[%
 reprint,twocolumn,longbibliography,preprintnumbers,
nofootinbib,
 amsmath,amssymb,w
 aps,
pra,
]{revtex4-2}
\pdfoutput=1
\usepackage[utf8]{inputenc}
\usepackage{flushend}
\usepackage{dcolumn}
\usepackage{bm}
\usepackage{balance}

\usepackage[normalem]{ulem}

\usepackage[colorlinks = true,
            linkcolor = blue,
            urlcolor  = blue,
            citecolor =green,
            anchorcolor = blue]{hyperref}
\usepackage{verbatim}
\usepackage{color,ulem}
\usepackage[english]{babel}

\usepackage{subcaption}
\usepackage{graphicx}

\usepackage[utf8]{inputenc}
\input Starburst.fd
\newcommand*\initfamily{\usefont{U}{Starburst}{xl}{n}}\initfamily

\newcommand{\beq}{\begin{eqnarray}}
\newcommand{\eeq}{\end{eqnarray}}
\usepackage{amsmath}
\usepackage{tikz}
\usetikzlibrary{decorations.pathmorphing}
\usetikzlibrary{shapes.misc}
\tikzset{cross/.style={cross out, draw=black, minimum size=8*(#1-\pgflinewidth), inner sep=0pt, outer sep=0pt},
cross/.default={1pt}}
\usetikzlibrary{patterns,math}
\begin{document}

\title{Hyperballistic transport in dense systems of charged particles under AC electric fields}

\author{Daniele Gamba$^{1,2}$, Bingyu Cui$^{3,4}$, Alessio Zaccone$^{1}$}
\email{alessio.zaccone@unimi.it}\email{daniele.gamba@theorie.physik.uni-goettingen.de}
\affiliation{${}^1$Department of Physics ``A. Pontremoli'', University of Milan, via Celoria 16,
20133 Milan, Italy}
\affiliation{${}^2$Institute of Theoretical Physics, University of G\"ottingen, Friedrich-Hund-Platz 1, 37077 G\"ottingen, Germany.}
\affiliation{${}^3$School of Science and Engineering, The Chinese University of Hong Kong, Shenzhen, Guangdong, 518172, P. R. China}
\affiliation{${}^4$Department of Chemistry, University of Pennsylvania, Philadelphia, Pennsylvania 19104,
USA}
\date{\today}

\begin{abstract}
The Langevin equation is ubiquitously employed to numerically simulate plasmas, colloids and electrolytes. However, the usual assumption of white noise becomes untenable when the system is subject to an external AC electric field. This is because the charged particles in the system, which provide the thermal bath for the particle transport, become themselves responsive to the AC field and the thermal noise is field-dependent and non-Markovian. 
We theoretically study the particle diffusivity in a Langevin transport model for a tagged charged particle immersed in a dense system of charged particles (plus also, possibly, other neutral particles) that act as the thermal bath, under an external AC electric field. This is done by properly accounting for the effects of the AC field on the thermal bath statistics.
We analytically derive the time-dependent generalized diffusivity $D(t)$ for different initial conditions. The generalized diffusivity exhibits damped oscillatory-like behaviour with initial very large peaks, where the generalized diffusion coefficient  is enhanced by orders of magnitude with respect to the infinite-time steady-state value. 
The latter coincides with the Stokes-Einstein diffusivity in the absence of external field. For initial conditions where the external field is already on at $t=0$ and the system is thermalized under DC conditions for $t \leq 0$, the short-time behaviour is hyperballistic, $MSD \sim t^4$ (where MSD is the mean-squared displacement), leading to giant enhancement of the particle transport. 
Finally, the theory elucidates the role of medium polarization on the local Lorentz field, and allows for estimates of the effective electric charge due to polarization by the surrounding charges.
\end{abstract}

\maketitle

\section{Introduction}

The Langevin equation has become one of the most widely employed mathematical tools to numerically simulate plasmas, colloidal systems and the molecular dynamics of liquids and solids \cite{Ceriotti}. The Langevin equation in its standard form reads as:
\begin{equation}
    m\frac{d\mathbf{v}}{dt}=-\gamma_0\mathbf{v}-\nabla V(\mathbf{x})+\mathbf{F}_p\left( t\right)
\label{eq:langevin1}\end{equation}
where $\mathbf{v}$ and $m$ are the particle's velocity and mass. respectively, $\gamma_0$ is a viscous friction coefficient, $V$ is a conservative field, and $\mathbf{F}_p$ is the stochastic force. The latter is typically assumed to obey the statistics of a white noise, as given by the (Markovian) fluctuation-dissipation theorem (FDT):
\begin{equation}
    \langle F_{i,p}\left( t\right)F_{j,p}\left( t'\right) \rangle =2\gamma_0 k_{B}T\delta _{i,j} \delta \left(t-t'\right)  \label{delta}
\end{equation}
where $i,j$ denote Cartesian components and $\delta \left(t-t'\right)$ is the Dirac delta function.
 
Ivlev and co-workers \cite{IvZhKl05} proposed a computational scheme to simulate plasmas and dusty plasmas via the Langevin equation, with a delta-correlated noise as in the above Eq.~\eqref{delta}.
Subsequent studies in the area of plasmas and dusty plasmas have adopted the same scheme \cite{Mabey2017,Graziani,Me19}, including 
dusty plasmas in oscillating (AC) external fields \cite{MaAkMa23,CoMoNo18,KaAlLi22,Rotenberg}, lane and band formation in dusty plasmas under AC fields \cite{VaIsSi22,VivaIm11,SuWyIv09}, and machine learning for searching phase transitions in dusty plasmas \cite{HuNoHu22}.

In most experiments, dusty plasmas are weakly ionized (ionization fraction around $10^{-6}$) and the thermal bath is almost a neutral gas. However, there are some theoretical studies that focus on complex, fully ionized plasmas \cite{Hansen1982, Hagstrom2011, Dan2017, Jia2016}: the thermal bath consists of charged particles and responds to the AC electric field \cite{CuiZaccone2018} (see also \cite{Thorwart_2016,Thorwart_2018} for the concept of externally-driven bath).
Other applications of the driven bath model lie in the field of electrolyte solutions subjected to AC electric fields \cite{vanRoij,vanRoij2,PhysRevLett.130.098001} and electrochemical devices \cite{Rotenberg}.

In Ref. \cite{CuiZaccone2018}, it was mathematically proved that, for systems of charged particles in an external AC electric field, the (generalized) Langevin equation is intrinsically non-Markovian. This is reflected in the fact that the fluctuation-dissipation theorem (FDT, i.e. the statistics of the thermal noise) is given (in 1D) by 
\begin{equation}
    \langle F_p(t)F_p(t')\rangle =m k_BT K(t-t')+Q^2 E(t)E(t') \label{FDT Cui}
\end{equation}
where $E(t)$ represents the external AC electric field
and $Q$ is  a renormalized effective charge related to $F_P$ via $\langle F_P \rangle = Q E(t)$. In this model, the generalized Langevin equation of motion for the charged particle is derived from a microscopic Caldeira-Leggett particle-bath Hamiltonian, where, crucially, the bath is formed of charged harmonic oscillators which also respond to the external AC field. This is a trick to effectively decompose the thermal kicks on the tagged particle due to innumerable collisions by decomposing the corresponding thermal bath into the vibrational eigenmodes of the system.

The modified FDT has an extra term $\sim E(t)E(t')$ in addition to the usual term $K(t-t')$. While the latter can, under certain conditions, be reduced to a Dirac delta as discussed e.g. in \cite{Zwanzig2001}, the term $\sim E(t)E(t')$ obviously cannot be reduced to a Dirac delta under any conditions, because $E(t) \sim \sin(\Omega t)$ is a sinusoidal function. This effect, therefore, arises only in the presence of an external time-dependent field, whereas, if the external field is DC, the effect is not there and Eq.~\eqref{delta} remains valid.

In this article, we adopt the same Caldeira-Leggett particle-bath model in Res. \cite{CuiZaccone2018} to study the charged particle transport in dense systems of charged particles (plus some other neutral particles as well), under an external AC electric field that is possibly relevant to plasmas, colloidal systems, liquid metals and ferrofluids, and dense electrolytes. The correct FDT for these systems is always intrinsically non-Markovian and is given by Eq.~\eqref{FDT Cui}. This fact has been systematically overlooked in the literature on Langevin simulations of plasmas, colloids and electrolytes, where the standard Markovian FDT Eq.~\eqref{delta} is used even when the plasma is under an AC electric or electromagnetic field. In the model presented in the next section, the electric charges for the baths degrees of freedom are taken to be fixed. This is not always the case in reality, for example, in dusty plasmas, particle charges are not constant but can fluctuate due to the stochastic nature of particle charging. In the presence of electromagnetic fields, these fluctuations would introduce additional terms in the Langevin equations, similar to those explored in this study. For further details, refer to \cite{Vaulina1999, Ivlev2000, Morfill2002}.

In this way, upon setting suitable initial conditions, we analytically derive the time-dependent particle diffusivity $D(t)$, which we study as a function of the AC electric field parameters, in particular the field amplitude $E_0$ and frequency $\Omega$. 
The magnitude of the effective polarization charge is estimated, and a giant enhancement of the polarization field is found when
the field frequency approaches the characteristic vibration frequency of the charged particles forming the bath. In the low-frequency limit, that is, when $\Omega$ is much smaller than the characteristic frequencies of oscillation of the medium, predictions concerning the behaviour of the generalized diffusion coefficient $D(t)$ are obtained including the prediction of superdiffusive transport when the external field frequency becomes very small. In particular, for the case of initial conditions where the AC field is switched on at $t=0$, this short-time superdiffusive regime is simply ballistic. Instead, if the system has thermalized under DC conditions for times $t \leq 0$ and then the field becomes AC for $t>0$, a hyperballistic superdiffusive regime is predicted.
It is also established that $D(t \rightarrow \infty)$ coincides with the particle diffusivity in the absence of the electric field under all initial conditions, which is a non-trivial result.


\section{Model of driven particle-bath systems}
We consider the same model as in Ref.~\cite{CuiZaccone2018}, in which an external, time-dependent electric field, $E(t)$, is applied to a dense medium. The motion of the tagged particle of mass $m$ is embedded in a dense medium (thermal bath) of other particles. Both the tagged particle as well as the particles (oscillators) forming the bath are electrically charged and dynamically respond to the external AC electric field.
In order to effectively decompose the effect of the thermal bath onto the normal modes of the system, we use the Caldeira-Leggett trick of representing the thermal environment as fictitious harmonic oscillators with natural frequencies corresponding to the eigenmodes of the medium. In our theory, these fictitious harmonic oscillators are electrically charged, and thus are subject to the effect of the external electric field via the Lorentz force. Hence, the thermal bath responds to the external field and its statistics is modified by the external field, as it should be, in the physical reality. We shall neglect magnetic forces throughout. 

The Hamiltonian $H$ (in 1D) of the total system consists of the Hamiltonian of the tagged particle  
\begin{equation}
H_p=\frac{p^2}{2m}-q\,x\,E(t),
\end{equation}
where $p$ is the momentum, $x$ the position and $q$ the particle's charge, plus the Hamiltonian $H_b$ of the bath. The latter is effectively modeled by $N$ independent electrically-charged harmonic oscillators of coordinates $\{x_i,p_i\}$, with mass $m_i$, electric charge $q_i$, and oscillating at some eigenfrequency $\omega_i$ \cite{Zwanzig2001,Caldeira1983}. Since each oscillator $i$ carries a charge $q_i$, it thus feels the electric force due to the external AC field, and the bath as a whole is responsive to the AC field. 
Following the Caldeira-Leggett approach, the Hamiltonian of the bath oscillators can be written as \cite{CuiZaccone2018}
\begin{align}
H_b&=\frac{1}{2}\sum_{i=1}^N\left(\frac{p_i^2}{m_i}+m_i\omega_i^2\left(x_i-\frac{\nu_i^2}{\omega_i^2}x\right)^2-2q_ix_iE(t)\right)\notag\\
&= H_b^0-
E(t)\sum_{i=1}^Nq_ix_i,
\label{eq:bathHamiltonian1}
\end{align}
with
\begin{equation}
    H_b^0=\frac{1}{2}\sum_{i=1}^N\left(\frac{p_i^2}{m_i}+m_i\omega_i^2\left(x_i-\frac{\nu_i^2}{\omega_i^2}x\right)^2\right).
\end{equation}
Here, $\nu_i^2$ is a coupling strength.

\begin{figure}[t]
\begin{subfigure}[b]{0.45\textwidth}
\includegraphics[width=0.9\linewidth]{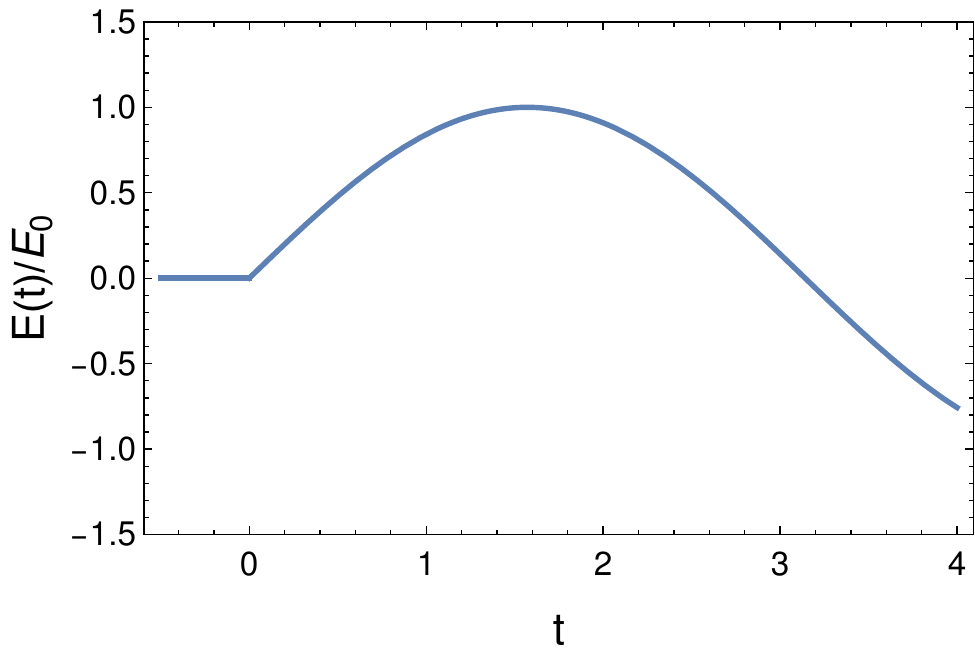}
   \caption{}
\end{subfigure}
\begin{subfigure}[b]{0.45\textwidth}
\includegraphics[width=0.9\linewidth]{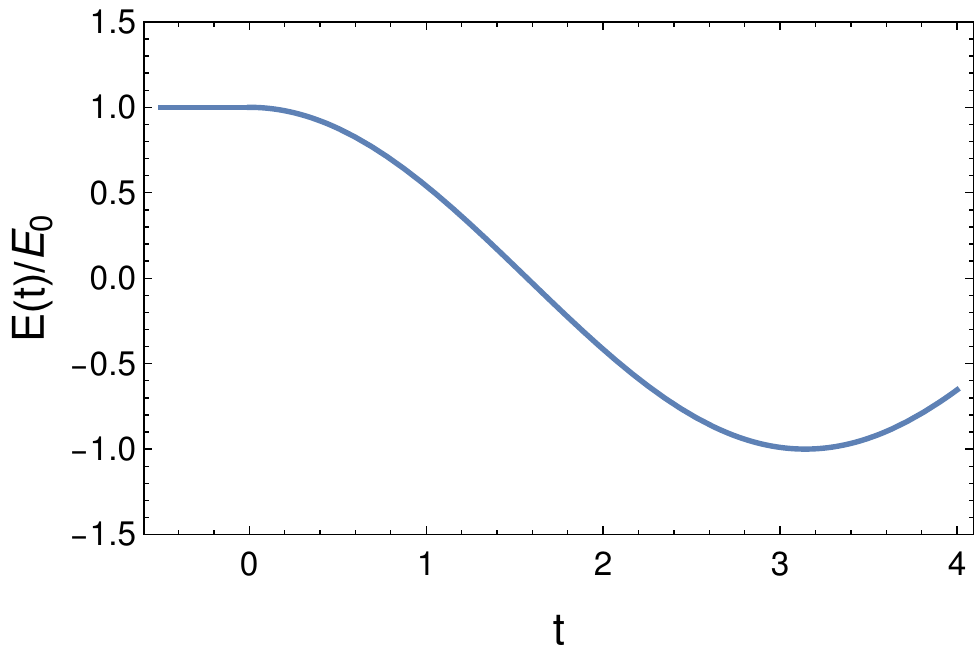}
   \caption{}
\end{subfigure}

\caption{Different initial conditions for the particle-bath Hamiltonian~\eqref{eq:bathHamiltonian1} under the action of an external AC electric field $E(t)$. (a): Field-off initial conditions, where the field is switched on at $t=0$ (Eq.~(\ref{eq:caseOne})). (b): Field-on initial conditions, where the field is already on at $t=0$ (Eq.~(\ref{eq:caseTwo})).}
\label{fig:condizioni}
\end{figure}

The external electric field oscillates with frequency $\Omega$ and amplitude $E_0$. We consider two different initial conditions (ICs): \textit{field-off initial conditions}, given by \begin{eqnarray}E(t)=\begin{cases} 
      0,&\ t<0\\
      E_0\sin(\Omega t),&\ t\geq 0
\end{cases}\label{eq:caseOne}\end{eqnarray}
and \textit{field-on initial conditions}, given by
\begin{eqnarray}E(t)=\begin{cases} 
      E_0,&\ t<0\\
      E_0\cos(\Omega t),&\ t\geq 0
\end{cases}.\label{eq:caseTwo}\end{eqnarray}
The field behavior in the two cases is shown 
in Fig.~\ref{fig:condizioni}.

\begin{figure}[h]
\centering
\label{fig:zero}
\begin{subfigure}[b]{0.45\textwidth}
\includegraphics[width=0.9\linewidth]{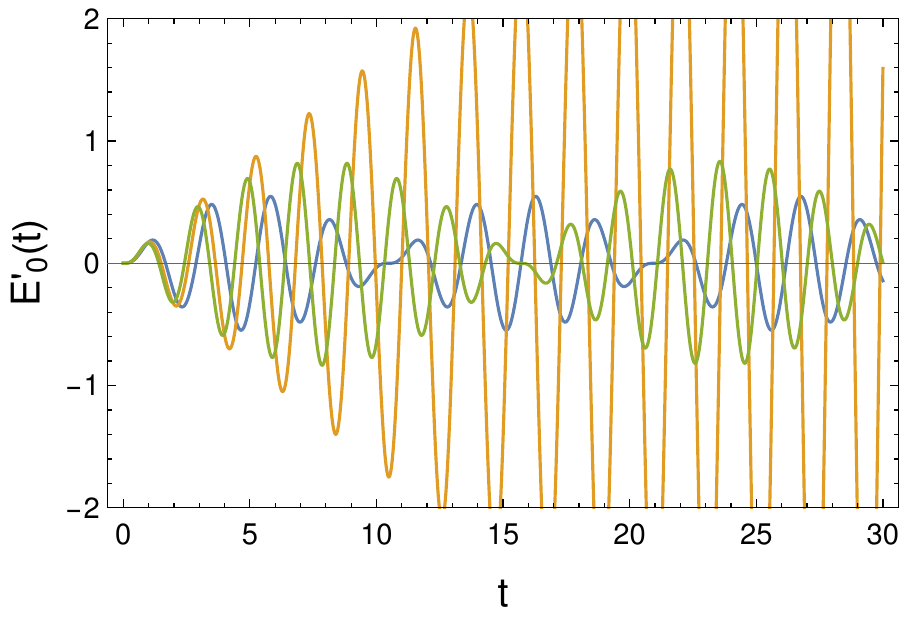}
   \caption{}
\end{subfigure}
\begin{subfigure}[b]{0.45\textwidth}
\includegraphics[width=0.9\linewidth]{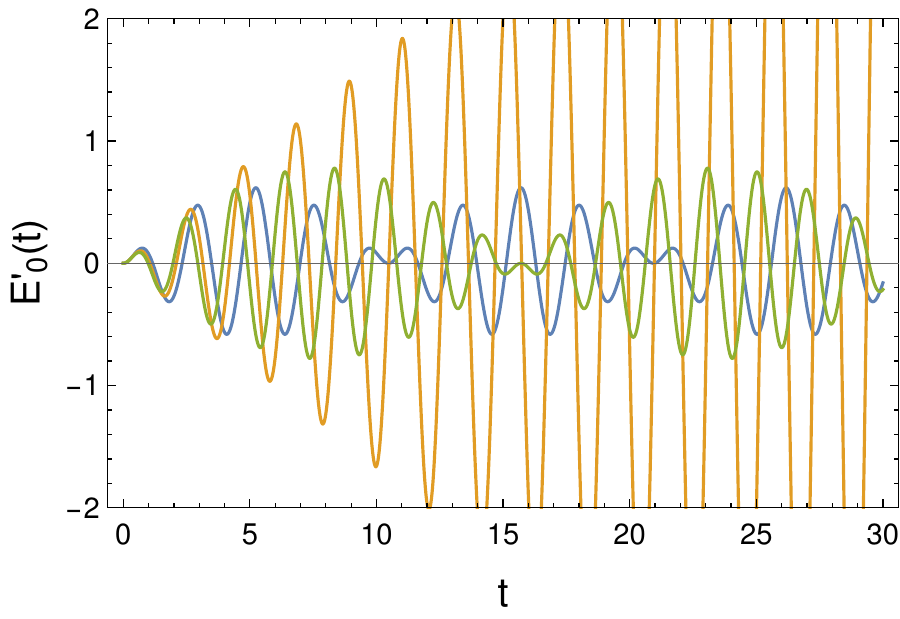}
   \caption{}
\end{subfigure}

\caption{Plot of the effective (Lorentz) field Eq.~\eqref{eq:campoEffettivo} (see also the more explicit form Eq.~\eqref{eq:lorentzFieldExplicit}, with $i=0$) as a function of time. Panel (a): field-off ICs. Panel (b): field-on ICs. Blue, yellow and green lines correspond to values $\Omega/\omega_0=0.8, 0.99, 1.13$.
The oscillations are giantly enhanced at the resonant frequency $\Omega\simeq\omega_0$, and their amplitude increases with time (Eq.~\eqref{eq:campoEffettivo}).}
\label{fig:E(t)}
\end{figure}

Following the canonical formalism, the derivation of Ref. \cite{CuiZaccone2018} gives the equation of motion of the tagged particle as
\begin{equation}
\frac{dp}{dt}=-\int_0^tK(t')p(t-t')dt'-\nabla V(x)+qE(t)+F_p(t),
\label{eq:generalizedLangevin}
\end{equation}
where 
\begin{equation}
K(t)=\sum_{i=1}^N\frac{m_i\nu_i^4}{m\omega_i^2}\cos(\omega_it)
\label{eq:memoryFunction}
\end{equation}
is the memory kernel describing the time accumulation effect on the tagged particle due to the coupling to the bath, and 
\begin{eqnarray}
F_p(t)&=&\sum_{i=0}^N\Bigg[m_i\nu_i^2\left(x_i(0)-\frac{\nu_i^2}{\omega_i^2}x(0)\right)\cos(\omega_it)
+\nonumber\\&&\ +\frac{\nu_i^2}{\omega_i}p_i(0)\sin(\omega_it)+q_iE'_i(\omega_i, t)\Bigg]
\label{eq:deterministicNoise}
\end{eqnarray}
where
\begin{eqnarray}
E'_i(\omega_i,t)=\frac{\nu_i^2}{\omega_i}\int_0^tE(t')\sin(\omega_i(t-t'))dt'.\label{eq:campoEffettivo}\end{eqnarray}  The ``noise" (\ref{eq:deterministicNoise}) depends on the distribution of the bath oscillators and on the electric field, and is a well-defined function of time. It describes the effect of a huge number of irregular kicks on the tagged particle due to the collisions with the particles of the bath. The force $\sum_{i=1}^Nq_iE'_i(t)$ acting on the Brownian particle is due to the internal polarization of the medium under the external AC field $E(t)$. The function $E_0'(\omega_0,t)$ (Eq. \eqref{eq:campoEffettivo}) is plotted in Fig. \ref{fig:E(t)} as a function of time.

In most cases, the medium contains a large number of oscillating modes, so we can regard the collection of oscillating eigenfrequencies as continuous. Therefore, the sum might be replaced with the integral $\int g(\omega)d\omega$, where $g(\omega)=\sum_i\delta(\omega-\omega_i)$ is the density of vibrational states. Thus, in the continuum limit, the friction kernel (\ref{eq:memoryFunction}) may be written as
\begin{equation}
K(t)=\frac{m_0}{m}\int_0^{\omega_D}\frac{\nu(\omega)^4}{\omega^2}\cos(\omega t)g(\omega)d\omega,\label{eq:memoria13}
\end{equation}
where we have set $m_i=m_0$ for all bath oscillators $i$. Following Zwanzig \cite{Zwanzig2001}, we assume a Debye form for the density of states of the bath:
\begin{equation}g(\omega)=
 \begin{cases} 
      g_0\omega^2,&  \ 0<\omega<\omega_D,\\
      0,&\ \text{otherwise,}
   \end{cases}
\end{equation}
where $g_0$ is a constant, and $\omega_D$ is the cut-off frequency (highest frequency) of the system (e.g. in a plasma this could be the electron plasma frequency). This choice is motivated by the fact that the main low-energy excitations in the system are sound waves with some dispersion relation $\omega(k) = c k$, where $c$ is the sound speed. The corresponding density of states of these low-energy eigenmodes $\omega$ is thus given by the quadratic Debye form \cite{Khrapak}. While also higher-energy vibration modes are potentially important in $g(\omega)$ (e.g. Langmuir waves etc), it should be noted that the factor $\omega^2$ in the denominator of the integrand function in Eq. \eqref{eq:memoria13} gives a larger weight to the low-energy vibration modes contained in $g(\omega)$. For a realistic plasma or dusty plasma, these will be vibration eigenmodes linked to the ion dynamics. For very dense dusty plasmas, they may coincide with acoustic modes of the dusty particles.

Equation \eqref{eq:memoria13} becomes \begin{eqnarray}
K(t)=g_0\frac{m_0}{m}\int_0^{\omega_D}\nu(\omega)^4\cos(\omega t)d\omega\label{eq:memoriaquindici}\end{eqnarray}
We also introduce, for later convenience, the renormalized effective charge
\begin{eqnarray}
Q=\sum_{i=1}^N\frac{q_i\nu_i^2}{\omega_i^2}=g_0q_0\int_0^{\omega_D}\nu(\omega)^2d\omega\equiv g_0q_0f(0),
\label{eq:caricaPolarizzazione}\end{eqnarray}
where we have set $q_i=q_0$ for all oscillators $i$, and
\begin{equation}
f(t)=\int_0^{\omega_D}\nu(\omega)^2\cos(\omega t)d\omega.
\end{equation}
The average noise $F_p$ and its autocorrelation function can be computed explicitly in the two situations described below.

\begin{figure}[h]
\centering

\begin{subfigure}[b]{0.45\textwidth}
\includegraphics[width=0.9\linewidth]{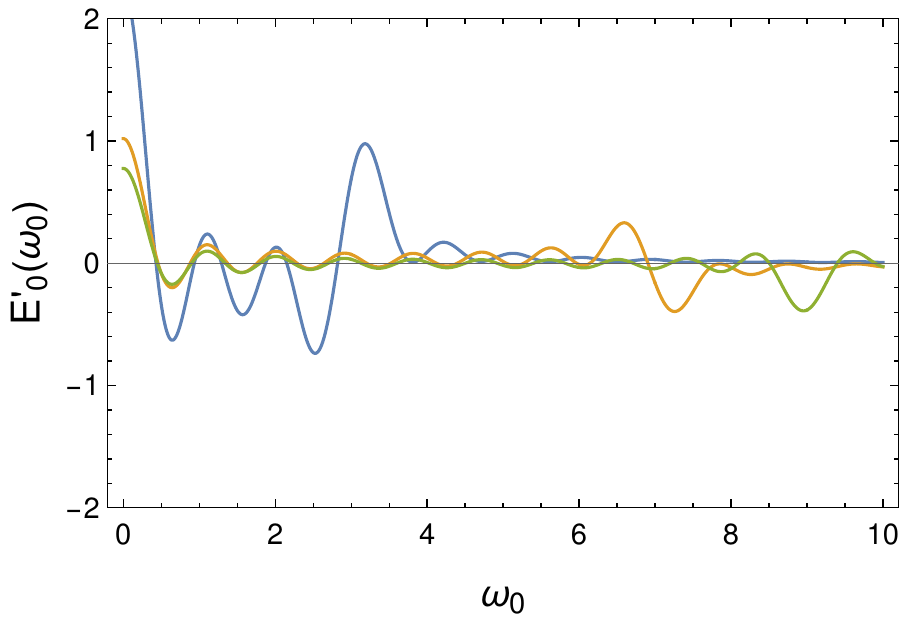}
   \caption{}
\end{subfigure}
\begin{subfigure}[b]{0.45\textwidth}
\includegraphics[width=0.9\linewidth]{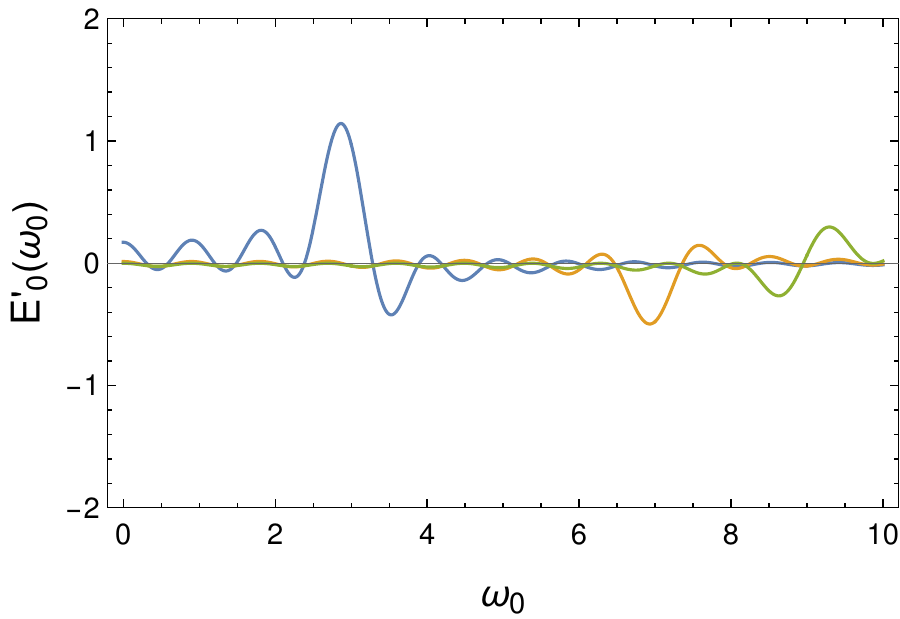}
   \caption{}
   \label{}
\end{subfigure}
\label{fig:zero-Uno}\caption{Plot of the maximum (with respect to $\Omega$) of the effective (Lorentz) field ~\eqref{eq:campoEffettivo} (see also the more explicit form~\eqref{eq:lorentzFieldExplicit}, with $i=0$) as a function of the dimensionless time $\omega_0 t$. 
Panel (a): field-off ICs. Panel (b): field-on ICs.  The maximum of $E_0'(\omega_0 t)$ is reached when the driving frequency $\Omega\sim\omega_0$. This maximum is rapidly increasing in time.}
\label{fig:E(w)}
\end{figure}

\subsection{Pseudo-resonance of the polarization function}
Consider the situation described by Eq.~(\ref{eq:caseOne}), where the external electric field $E$ is switched on at time $t=0$. However, for very short times, the bath doesn't have time to instantaneously respond to the change, and is still in the same equilibrium distribution $\sim e^{-H_b^0/k_BT}$ as at $t=0$ and earlier times. Therefore, the average noise becomes 
\begin{equation}
\langle F_p(t)\rangle=
\sum_{i=0}^N\frac{q_i\nu_i^2E_0}{\omega_i}\frac{\omega_i\sin(\Omega t)-\Omega\sin(\omega_it)}{\omega_i^2-\Omega^2}.
\label{eq:dieciCoseni}
\end{equation}
For the large values of $\Omega$, that are used for instance in THz spectroscopy, it can become comparable with the smallest of the $\omega_i$ values, say $\omega_0$, then
\begin{equation}
\langle F_p(t)\rangle\sim q_0E'_0(\omega_0,t)=\frac{q_0\nu_0^2E_0}{2\omega_0}\left[\frac{\sin(\omega_0 t)}{\omega_0}-t\cos(\omega_0 t)\right]
\label{eq:undiciCosines}\end{equation}
and the tagged particle is subject to an oscillating force whose amplitude grows linearly in time (see Section~\ref{app:1} in the Appendix for the detailed derivation of Eq.~\eqref{eq:dieciCoseni} and~\eqref{eq:undiciCosines}). It should be noted that, in a plasma or dusty plasma, the characteristic frequency $\omega_0$ may coincide with a low-energy acoustic mode of the ions. For very dense dusty plasmas, this lowest characteristic frequency may coincide with an acoustic mode of the dust particles.

The trend of $E'_0(\omega_0,t)$ (Eq.~\eqref{eq:campoEffettivo}) is shown in Fig. \ref{fig:E(t)}, for both the field-off and the field-on ICs. This figure shows that when $\Omega\sim\omega_0$, the Lorentz field undergoes enormous amplification in time. Figure~\ref{fig:E(w)} shows how the maximum (with respect to $\Omega$) of the Lorentz field, which is reached for $\Omega\sim\omega_0$, grows in time.

Now we consider the situation described by Eq.~(\ref{eq:caseTwo}) in which at $t=0$, the electric field has already been on for a time sufficiently long for the bath to equilibrate under the action of the external field. The usual relaxation time for atomic/molecular systems is of order $10^{-13}$ seconds, so possibly much shorter then the time of observation. The state of the medium is therefore thermalized according to the distribution $\sim e^{-H_b(t=0)/k_BT}$, with which we find
\begin{equation}
\langle F_p(t)\rangle=\sum_{i=1}^Nq_i\nu_i^2E_0\Bigg\{\frac{\cos(\omega_it)}{\omega_i^2}+
\frac{\cos(\Omega t)-\cos(\omega_it)}{\omega_i^2-\Omega^2}\Bigg\}.
\label{eq:mediaGeneralizzata2}\end{equation}

Again, if $\Omega$ is comparable with some resonant frequency $\omega_0$, then 
\begin{equation}
\langle F_p(t)\rangle\sim q_0E'_0(\omega,t)=\frac{q_0\nu_0^2E_0}{2}\frac{\sin(\omega_0 t)}{\omega_0}t,
\label{eq:linearTime}\end{equation}
and we have an oscillating force whose amplitude increases linearly in time
(see Section~\ref{app:1} in the Appendix for the detailed derivation of Eq.~\eqref{eq:mediaGeneralizzata2} and~\eqref{eq:linearTime}).

As we see, Eqs.~(\ref{eq:dieciCoseni},~\ref{eq:mediaGeneralizzata2}) exhibit a pole in coincidence of 
$\Omega\simeq\omega_0$. As explained in \cite{Jackson1998}, this pole represents a resonance peak, when the experimental frequency $\Omega$ is probing exactly a characteristic frequency $\omega_0$ of the material medium, and this gives rise to spectra of emission and absorption. One usually solves the problem of integrating the singularity, by displacing the pole by a little amount in the complex plane: this physically corresponds to recognizing that each mode $\omega_i$ also suffers some dissipation. In that case however, the resonance is made possible when the probing oscillating field has been on for a long amount of time. In our case, the divergence is avoided because we have been probing the medium on its resonance frequency only for a time $t$. 

These frequency sums can be evaluated analytically. One should note that the sum contains diverging quantities, which, however, when summed together give rise to a finite quantity, due to a negative interference effect. The result is a quantity that is oscillating in time, and the giant amplitude enhancement that grows linear in $t$ has no effect on the quantity integrated in time, because all the giant oscillating terms average to quantities of order 1. Interesting effects might appear if we consider the probing frequency $\Omega$ comparable with the highest eigenfrequency $\omega_D$ of the medium, but that goes beyond the scope of the present paper and may be the subject for future work.

\begin{figure}[h]
\centering
\centering
\includegraphics[width=0.9\linewidth]{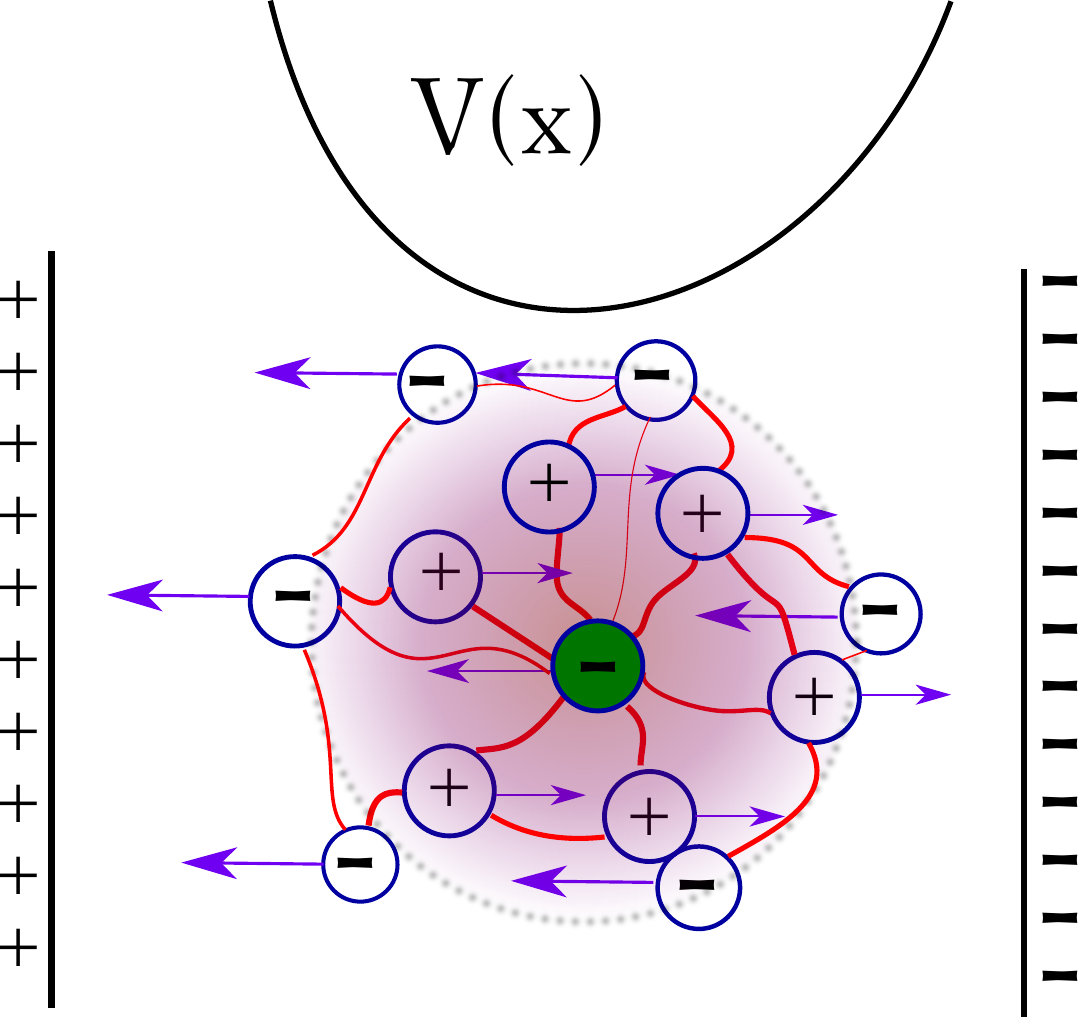}
\caption{Schematic depiction of the physical system (dense ionized matter) considered in the theoretical model. The tagged particle (green) moves in a thermal bath (dense medium) of other charges schematically represented as charged harmonic oscillators in the theoretical model. The frequencies of these fictitious oscillators schematically represent the eigenfrequencies of the sound modes of the system. These oscillators are coupled to the tagged particle by different coupling strengths, represented as red lines. The effective charge of the tagged particle, $Q$ (represented as a shaded area), accounts for the local charge distribution and polarizing Lorentz field around the tagged particle. The system is under an AC electric field (depicted in figure at a certain instant of time) exerted by the electrodes (represented by two opposite charged plates in the graph) and might be subjected to an overall static field $V(x)$, which is not explicitly taken into account in our derivations but the presence of which would not qualitatively change our results. }
\label{fig:Tre}
\end{figure}

In radiofrequency (RF) plasma devices operating at typical AC frequencies of $13$ MHz, electrons can effectively respond to the alternating electric field, whereas ions cannot. The frequencies of the harmonic oscillators constituting the bath correspond with the electron and ion plasma frequencies. Depending on the plasma type, the electron plasma frequency can vary widely, ranging from $10^9$ Hz (gas discharge) to $10^{13}$ Hz (fusion machines).
Therefore, in scenarios like THz spectroscopy, where very high values of $\Omega$ are utilized, it is possible for $\Omega$ to approach or even surpass the smallest frequency $\omega_i$ of the bath, denoted here as $\omega_0$. Notably, in certain fusion devices, THz frequencies could correspond to the smallest $\omega_i$ values.
Having made the above considerations, we can keep working in the limit $\Omega\ll\omega_i$ for every mode $i$. 
The following theory is therefore developed in the limit of low driving frequency of the external AC field, i.e. for $\Omega\ll 10^9\text{Hz}$.

\subsection{Limit of the theory for low frequencies}
In this limit the sum in Eq.~\eqref{eq:dieciCoseni} can be evaluated and gives
\begin{eqnarray}
\langle F_p(t)\rangle=QE_0\sin(\Omega t)\label{eq:venti}\end{eqnarray}
for field-off ICs; similarly Eq.~\eqref{eq:mediaGeneralizzata2} gives
\begin{eqnarray}
\langle F_p(t)\rangle=QE_0\cos(\Omega t)\label{eq:ventuno}\end{eqnarray}
for field-on ICs, 
where \begin{eqnarray}
Q=g_0q_0\nu_0^2\omega_D\label{eq:caricaVentitre}
\end{eqnarray} (see App.~\ref{app:1} for the detailed derivation of Eqs.~(\ref{eq:venti},~\ref{eq:ventuno},~\ref{eq:caricaVentitre})).
So, as we can see, the quantity $Q$ has the role of an effective or renormalized charge acquired by the tagged particle, due to the polarization of the surrounding medium. The physical picture is that, in dense or supercooled liquid ionized matter, each particle is trapped in a cage by the surrounding particles. If the oscillatory AC field drives the particle in one direction and opposite charges in the opposite direction, then we will have a resonance effect with the normal mode of the particle in the cage. The effect adds up to the electric field on the particle, because in general the positive ions draw the particle in the opposite direction of the field, and if $q=-Q$ they would completely screen the effect of the external AC field. See Fig.~\ref{fig:Tre}
for a schematic visualization of this effect. It is also important to note that results~(\ref{eq:venti},~\ref{eq:ventuno}) show insensitivity of the system to the choice of initial conditions, i.e. whether field-off or field-on, because memory of the initial conditions is lost in a time proportional to $\omega_D^{-1}$, which is a very short time, e.g. if $\omega_D$ coincides with the electron plasma frequency. Indeed, in (\ref{eq:mediaGeneralizzata2}), in the limit $\Omega\ll\omega_D$, the first term, integrated on the frequency, is proportional to $\frac{\sin(\omega_D t)}{t}$, becomes negligible in comparison with the second term, proportional to $\omega_D$, because the time of observation is $t\gg\omega_D^{-1}$. Hence, the extra term due to field-on initial conditions becomes negligible after very short times, and the dependence on the initial conditions is lost.


\subsection{Autocorrelation functions}
The influence of the external field is also exhibited in the autocorrelation function of the random force,
\begin{align}
\langle F_p(t)F_p(t')\rangle=mk_BTK(t-t')+(QE_0)^2\sin(\Omega t)\sin(\Omega t'),
\label{eq:teoremaFluttuazioneZaccone}
\end{align}
as has been proposed in \cite{CuiZaccone2018}, for field-off initial conditions, and
\begin{align}
\langle F_p(t)F_p(t')\rangle&=mk_BTK(t-t')+(QE_0)^2\cos(\Omega t)\cos(\Omega t')
\label{eq:teoremaFluttuazioneGenerale}
\end{align} for field-on initial conditions (here derived for the first time;
the  details of the
calculation
are presented in Appendix ~\ref{app:1}).
In contrast to the previous case of the field-off ICs, the effective Lorentz field arising inside the medium is phase-shifted according to the external electric field. 

We can also calculate the momentum autocorrelation from the force correlation in the overdamped Brownian limit, i.e., when $\gamma_0\gg m$ (see Eq. \eqref{eq:langevin1}). 
Setting $\nu(\omega)=\nu_0$, Eq. \eqref{eq:memoriaquindici} becomes 
$K(t)=\gamma_0\delta(t)$
with 
\begin{equation}\gamma_0=\pi g_0\frac{m_0}{m}\nu_0^4.
\end{equation}
We find
\begin{equation}
\langle p(t_1)p(t_2)\rangle=\frac{1}{\gamma_0^2}\langle F(t_1)F(t_2)\rangle,
\label{eq:VACF}
\end{equation} 
with
\begin{eqnarray}
F(t)=F_p(t)+qE(t).
\end{eqnarray}

\section{Particle diffusivity under external AC field}
Using the time correlation of the velocity $v=p/m$, we are able to calculate the diffusivity
\begin{equation}
D(t)=\frac{1}{2t}\langle\{x(t)-x_0\}^2\rangle,
\end{equation}
where $x_0$ is the position of the Brownian particle at $t=0$, since
\begin{equation}
\langle\{x(t)-x_0\}^2\rangle=\int_0^tdt_1\int_0^tdt_2\langle v(t_1)v(t_2)\rangle.
\label{eq:eqMSDm}
\end{equation}

For the field-off ICs, the diffusivity becomes (see Appendix B):
\begin{equation}
D(t)=
\frac{k_BT}{m\gamma_0}\left(1+\mathcal{E}_0^2\frac{(\cos(\Omega t)-1)^2}{\Omega t}\right)
\label{eq:diffusivity1}
\end{equation}
with 
\begin{equation}
\mathcal{E}_0=\frac{(q+Q)E_0}{\sqrt{mk_BT\gamma_0\Omega}}
\label{eq:E1},
\end{equation}
while for the field-on ICs, we find
\begin{equation}
D(t)=\frac{k_BT}{m\gamma_0}\left(1+\mathcal{E}_0^2\frac{\sin^2(\Omega t)}{\Omega t}\right)
\label{eq:diffusivity2}
\end{equation}
(see App.~\ref{app:2} for a detailed derivation). 

Interestingly, we shall note that both the above expressions for $D(t)$ asymptotically recover the particle diffusivity in the absence of the external field in the limit $t\rightarrow \infty$:
\begin{equation}
    \lim_{t\rightarrow \infty}D(t) =\frac{k_BT}{m\gamma_0},\label{Stokes}
\end{equation}
which is the standard Stokes-Einstein result, independent of the AC field parameters.
Recalling the discussion following Eq.~\eqref{eq:linearTime}, the generalized diffusivity $D(t)$ doesn't show a divergent behavior for frequencies near the resonant frequencies of the system: in fact, in Eqs.~(\ref{eq:diffusivity1},~\ref{eq:diffusivity2}), the resonant frequencies of the system affect only the parameter $\mathcal{E}_0$, trough the renormalized charge $Q$.

\begin{figure}[t]
\centering
\centering
\begin{subfigure}[b]{0.5\textwidth}
\includegraphics[width=0.9\linewidth]{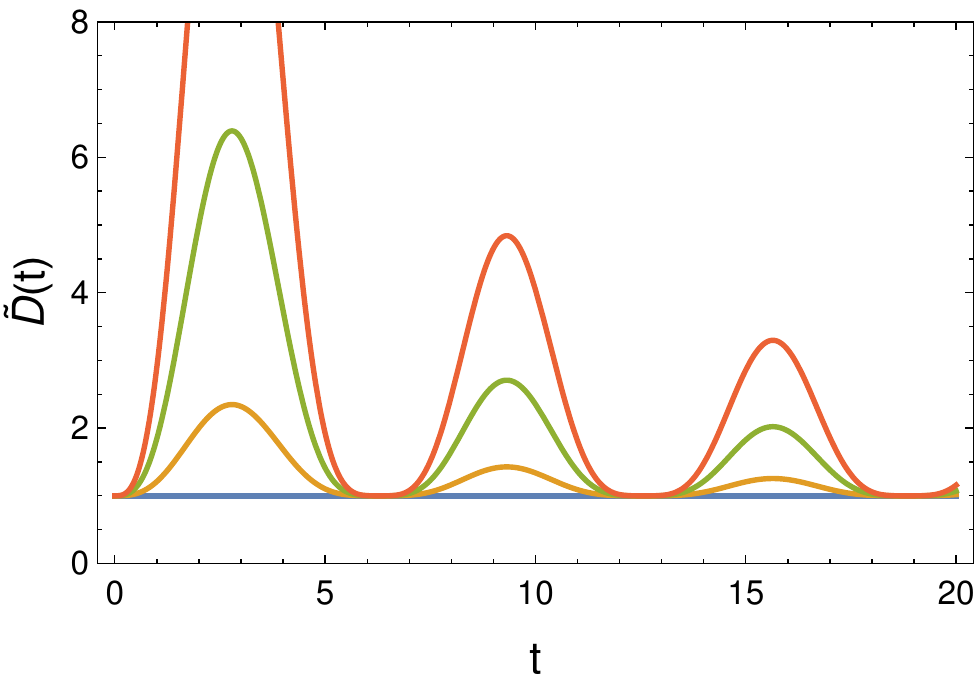}
   \caption{}
\end{subfigure}
\begin{subfigure}[b]{0.5\textwidth}
\includegraphics[width=0.9\linewidth]{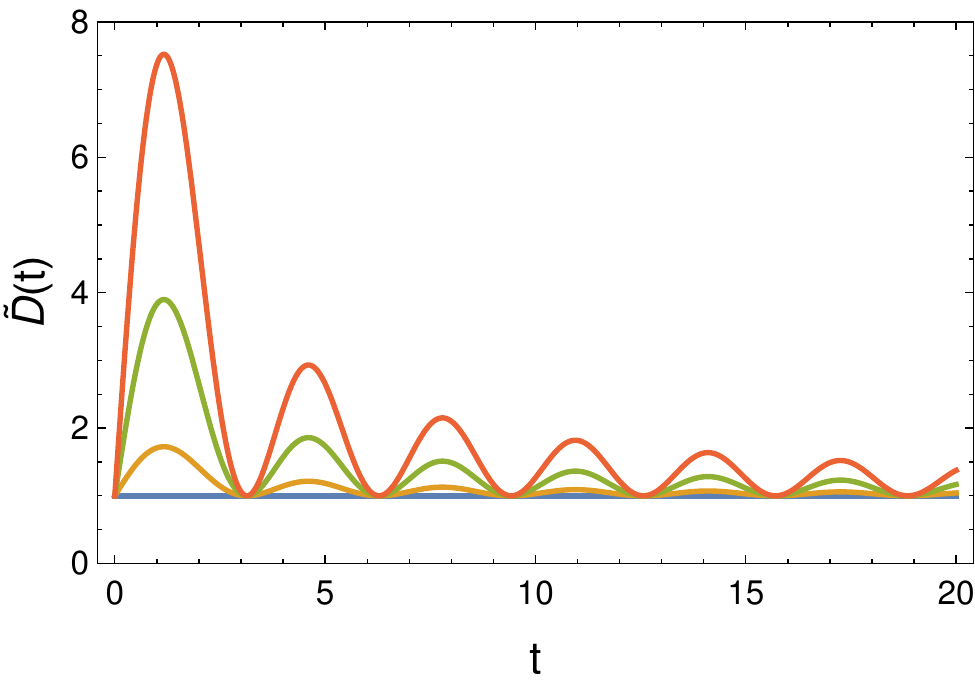}
   \caption{}
   \label{fig:Ng2-1}
\end{subfigure}
\caption{
 Time evolution of the generalized diffusivity $\tilde{D}=D\frac{m\gamma_0}{k_BT}$ for (a) field-off~\eqref{eq:diffusivity1} and (b) field-on~\eqref{eq:diffusivity2} initial conditions, obtained with different field amplitudes $\mathcal{E}_0$ (Eq.~\eqref{eq:E1}). 
 Yellow, green and orange lines correspond to $\mathcal{E}_0=1,2,3$, whereas the blue line is the Stokes-Einstein result obtained for $\mathcal{E}_0=0$. In the inset, the red line shows the diffusivity's behavior for small times. For comparison, the green line in the inset of panel (a) is $\sim t^3$, while the green line in the inset of panel (b) is $\sim t$.
}
\label{fig:1}
\end{figure}

\begin{figure}[h]
\centering
\begin{subfigure}[b]{0.5\textwidth}
\includegraphics[width=0.9\linewidth]{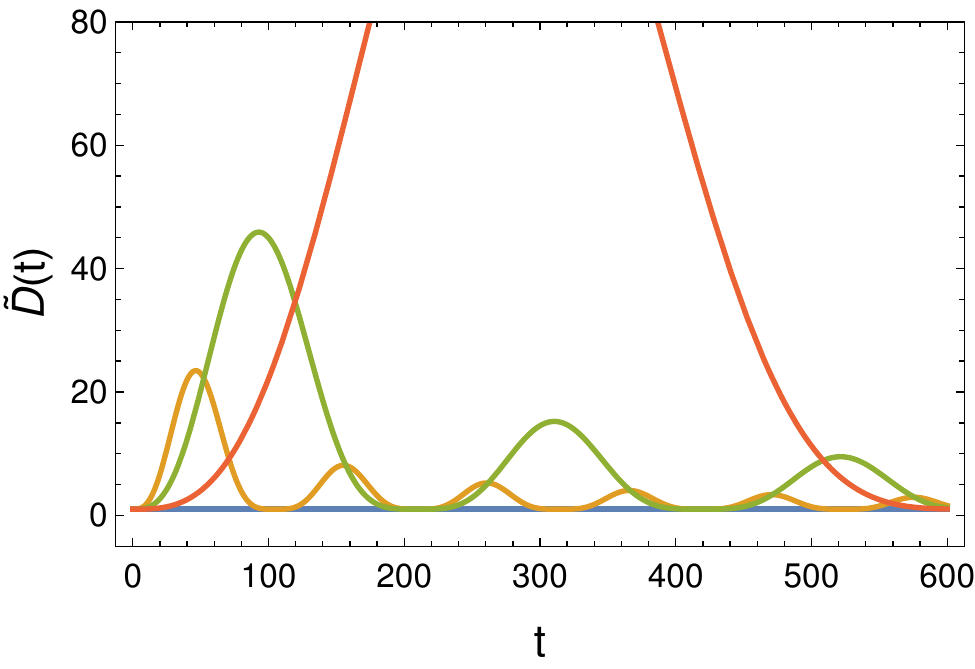}
   \caption{}
   \label{fig:Ng1} 
\end{subfigure}
\begin{subfigure}[b]{0.5\textwidth}
\includegraphics[width=0.9\linewidth]{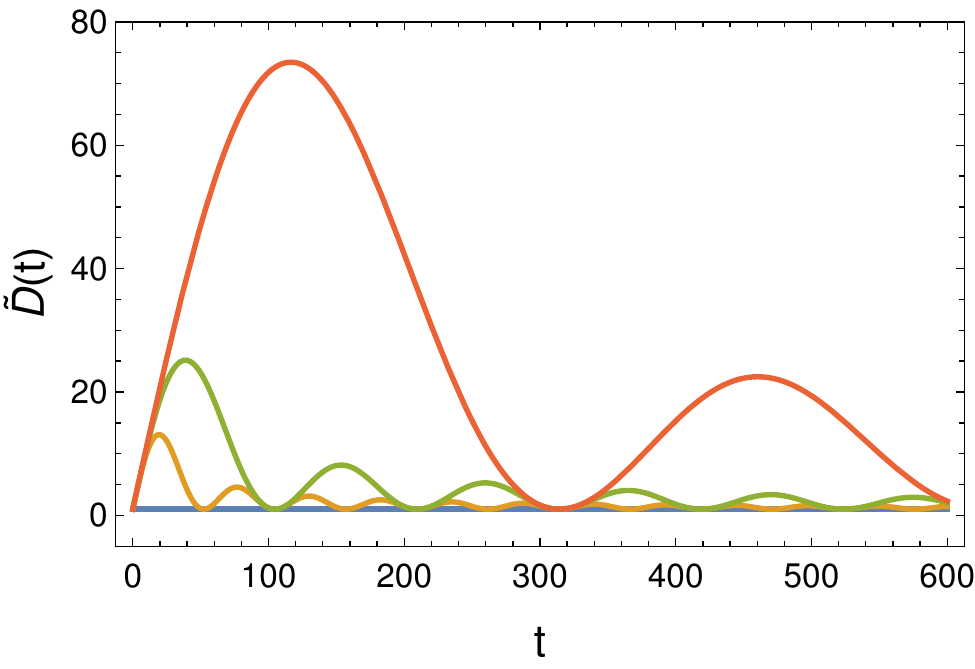}
   \caption{}
   \label{fig:Ng2}
\end{subfigure}
\caption{Time evolution of the generalized diffusivity $\tilde{D}=D\frac{2m\gamma_0}{k_BT}$ for (a) field-off~\eqref{eq:diffusivity1} and (b) field-on~\eqref{eq:diffusivity2} initial conditions plotted on a common time scale. 
Yellow, green and orange lines correspond to $\mathcal{E}_0^2=2.5,~5,~10$,  whereas the blue line is the Einstein result obtained for $\mathcal{E}_0=0$. 
For small times the diffusivity is giantly augmented in comparison with the DC diffusivity, which coincides with the Stokes-Einstein result (blue line).
}

\label{fig:2}
\end{figure}

The non-dimensionalized diffusivity $\tilde{D}(t)=\frac{m\gamma_0}{k_BT}D(t)$ associated with the motion of the tagged particle under the field-off ICs is shown in Figs. \ref{fig:1} and \ref{fig:2}.
In Fig. \ref{fig:1}, we compare $D(t)$ for  different values of the non-dimensional field amplitude Eq. \eqref{eq:E1}. In Fig. \ref{fig:2}, we compare $D(t)$ for different values of the non-dimensional amplitude $\mathcal{E}_0^2$.

Overall, the oscillatory diffusivity ultimately decays towards the $t\rightarrow\infty$  Stokes-Einsten value~\eqref{Stokes} as if there was no field applied. The effect of the AC field under the field-off ICs is more significant at short times, resulting in a first peak of $D(t)$ which is larger for larger values of $E_0$. 
Similar behaviour is also reported for the response of a colloidal particle to a time-dependent quadratic potential due to an optical trap being dragged through the fluid \cite{Ritort2020}.  In our case, these oscillations are augmented by a factor $Q$ due to the polarization of the surrounding medium (Eq.~(\ref{eq:E1})). If the external frequency~$\Omega$ is 
very small, in the case of field-on ICs we recover for times $t<\Omega^{-1}$ the limit of ballistic motion under a constant field, that is, the limit for small $\Omega$ of (\ref{eq:diffusivity2}), which is
\begin{equation}
D(t)\approx \frac{k_BT}{m\gamma_0}\left(1+\mathcal{E}_0^2\Omega t\right)
\label{eq:diffusivity2_approx}
\end{equation}
This asymptotic behavior for small $t$ is visible in the initial slope in Figs. \ref{fig:Ng2-1} and \ref{fig:Ng2}.

Equation \eqref{eq:diffusivity2} shows that the first peak in $D(t)$ is at $\Omega t=1.16$: \begin{eqnarray}
D_{\text{max}}\simeq D(1.16
/\Omega)=\frac{k_BT}{2m\gamma_0}\left(1+0.72\mathcal{E}_0^2\right).\end{eqnarray}
Since $\mathcal{E}_0^2\propto 1/\Omega$, when $\Omega$ is decreased, the peak is shifted forward in time and amplified.
In the DC limit of constant field, i.e. for small $\Omega$, and for finite times, Eq.~\eqref{eq:diffusivity2_approx} reduces to the Stokes-Einstein value, plus a standard drift contribution, linear in time, which contributes to the \textit{dressed} or generalized diffusivity $D(t)$, while the \textit{naked} diffusivity is given simply by the Stokes-Einstein value.

In the case of field-off ICs, instead, the short-time behaviour is given by taking the limit of small $\Omega t$ in Eq.~\eqref{eq:diffusivity1}, which gives:
\begin{equation}
D(t)\approx \frac{k_BT}{m\gamma_0}\left(1+\frac{1}{4}\mathcal{E}_0^2\Omega^3 t^3\right)
\label{eq:diffusivity1_approx}
\end{equation}
and, in this case, there is an inflection point at $t=0$. Importantly, in this case, a hyperballistic behaviour is observed at short times, $D(t) \sim t^3$, giving a faster and larger growth of the first peak in $D(t)$ compared to the previous case. This corresponds to a mean squared displacement (MSD) that grows as $\langle {x(t)-x_0}^2 \rangle \sim t^4$, i.e. with the fourth power of time. This behavior is shown in the inset of Fig.~\ref{fig:Ng2-1}.

This is a consequence of the AC field $\sim\sin(\Omega t)$ being switched on at $t=0$ following thermalization under DC conditions. In the overdamped regime, this gives a particle displacement $\cos(\Omega t)-1\sim t^2$.
Eq. \eqref{eq:diffusivity1} shows that the first peak of $D(t)$ is at $\Omega t=2.78$:
\begin{eqnarray}
D_{\text{max}}\simeq D(2.78/\Omega)=\frac{k_BT}{2m\gamma_0}\left(1+1.35\mathcal{E}_0^2\right).\label{eq:diffusivity1Approximated}\end{eqnarray}
When $\Omega$ is decreased, the peak is shifted forward in time and amplified.
In the $\Omega t\ll 1$ limit, Eq.~\eqref{eq:diffusivity1Approximated} shows that, for finite times, the Stokes-Einstein value (blue line in Fig. \ref{fig:Ng1}) is recovered.
This is the first prediction, to our knowledge, of hyperballistic superdiffusive transport in dense systems of charged particles under an AC electric field.

Consistent with results in Ref. \cite{Makabe1997}, while the charge is accelerating under the external field, large velocities are smeared out by the damping $\gamma_0$. Charges moving faster will be dragged by the friction to a larger extent. For a lump of charge density moving in the field direction, the charges in the front are faster and undergo major damping, while those on the rear do not undergo much frictional resistance and do not lose much momentum. As a result, charges accumulate in the front, increasing the local charge density. The overall effect is that the pulse gets compressed in the direction of the field, which reduces diffusivity parallel to the field. On the other hand, the transverse diffusivity might increase. In the example of diffusion-convection of a solute in a liquid, cfr. Ref. \cite{Probstein1989}, there is complete uniformity in Taylor dispersion transverse to the applied field, so there is no rise in gradients of velocity or concentration. However, in this paper, we are only considering the longitudinal case for the 1D Langevin equation along the field direction. The transverse equation might be coupled to what happens in the longitudinal one, but this would require a full 3D treatment which may not be analytically tractable within the Langevin equation framework.

As we have seen from the above theory, the correction to the (naked) diffusivity coefficient due to the external AC electric field is proportional to $\omega_D$ (this is because $\mathcal{E}_0 \propto Q \propto \omega_D$, see Eqs.~(\ref{eq:E1},~\ref{eq:caricaVentitre})). 
In the example of plasmas, since this is the highest frequency of the bath, it can be taken as the plasma frequency, and hence,\begin{eqnarray}
\omega_D\propto\frac{1}{\sqrt{m}}
\end{eqnarray} where $m$ is the mass of the electron, or in general, of the lightest charged particle in the medium.  
This implies that, in general, the diffusivity enhancement effect predicted by the above theory will be larger the smaller the mass of the lightest constituent present in the medium.

\section{Conclusions}
We studied a Caldeira-Leggett particle-bath model where both the tagged particle and the oscillators forming the bath are electrically charged and respond to an external AC electric field. Since the bath is responding to the external AC field, the thermal noise is no longer white noise, but is non-Markovian and strongly influenced by the external time-dependent electric field. 
Based on this model, we analytically derived results for the autocorrelation functions of force and momentum, and used the latter to evaluate the time-dependent generalized particle diffusivity $D(t)$. Analytical close-formed expressions for $D(t)$ as a function of the physical parameters of the system are obtained for two different initial conditions (AC field off and AC field on at $t=0$, respectively), and are given by Eq.~\eqref{eq:diffusivity1} and \eqref{eq:diffusivity2}.
We found that the generalized diffusivity in the asymptotic $t \rightarrow \infty$ steady state is the same as in the absence of the external field and is given by the Stokes-Einstein formula, which is a rather remarkable result. Also, the generalized diffusivity exhibits time-dependent damped fluctuations in the transient regime which, for low AC field frequency $\Omega$, produce a transient giant enhancement of the generalized diffusivity with respect to the steady-state value. For low enough frequency of the external AC field, the enhancement can be up to few orders of magnitude with respect to the steady-state value. 
In particular, our theory predicts: (i) \emph{ballistic superdiffusion} for the field-off initial conditions (the AC field is switched on at $t=0$ following equilibration at zero external field), and (ii) \emph{hyperballistic superdiffusion} with quartic power of the mean squared displacement, $MSD \sim t^4$, for the case of field-on initial conditions (the AC field is switch on at $t=0$ following thermalization under DC field).

This effect, predicted here for the first time, can be interpreted as the result of a memory accumulation (speed-up) process, due to the non-Markovianity of the noise  (Eqs. \eqref{eq:teoremaFluttuazioneZaccone} and \eqref{eq:teoremaFluttuazioneGenerale}). This is made evident if we take the DC limit of constant field ($\Omega \rightarrow 0$): in this case, both the naked diffusivities corresponding to \eqref{eq:diffusivity2_approx} and \eqref{eq:diffusivity1_approx} are reduced to the Stokes-Einstein value.
From a physical point of view, this memory accumulation means that the stochastic force acting on the particle, representing collisions with other particles, retains memory of previous collisions which in the field-on ICs case are continuously enhanced by the drift over past times, and hence grows as a function of time. A similar non-Markovian memory-accumulation effect can play a role in the speed-up of reaction kinetics as suggested in \cite{Florian} for the example of NaCl dissociation kinetics in water.
Also, the enhancement of the transient generalized diffusivity is predicted to be inversely proportional to the square root of the lightest particle in the medium (this could be e.g. the electron mass in the example of plasmas).

The theory also allows one to evaluate the local (Lorentz) field acting on the tagged particle due to polarization in the surrounding medium. Interestingly, a resonance-type behaviour of the local effective field is unveiled corresponding to a frequency value of the external AC field that matches the value of a pole in one of the terms constituting the average random force $\langle F_p(t) \rangle$ acting on the tagged particle. Also interestingly, however, a destructive interference with other oscillating terms in $\langle F_p(t) \rangle$ conspires to remove this resonance in the overall behaviour of $\langle F_p(t) \rangle$ which turns out to be given by the same sinusoidal function of the external AC field, with a renormalized charge due to charge renormalization by the surrounding charged particles.  

Since the diffusivity is a key parameter in plasma physics, e.g. it is a control parameter for the formation of plasma blobs in tokamaks \cite{Bisai2023,TheilerPRL} and plasma turbulence \cite{Ricci_2012}, the predicted effect may be useful in future applications for optimizing transport phenomena and the overall efficiency of fusion reactors.

From the fundamental point of view of nonequilibrium statistical mechanics, we have discovered and predicted a new physical effect which represents a new example of anomalous diffusion \cite{Metzler,Bouchaud1990}, namely of hyperballistic superdiffusion \cite{superuniv,Haenggi}. In particular, we showed that dynamic accumulation of the stochastic force, which represents random collisions enhanced by an external drift, leads to hyperballistic transport \cite{Baldelli}. Our theoretical findings may connect to previous experimental reports of hyperballistic transport in photons propagating through dynamic disorder~\cite{Levi2012,Peccianti2012}. 
The predicted effect may also bear some connection to the phenomenon of secondary Fermi acceleration, proposed to explain the origin and acceleration of cosmic rays \cite{Fermi}. This is the mechanism by which a plasma is heated and accelerated by time-dependent stochastic driving fields. In the original setup, this was a model for cosmic rays being scattered by magnetized interstellar clouds that move randomly and act as magnetic mirrors. 

However, we shall emphasize that there already exist well-developed transport theories for fusion plasmas \cite{Ho1987,Rudakov1971}, astrophysical plasmas \cite{Chaudhary2018} and high energy density matter \cite{PaulDrake2018}. While not yet directly applicable to practical plasmas, the purpose of our model is not to replace these other theories; it will instead serve as a complementary approach and as a starting point to provide new insights into the Langevin-based modelling of transport behaviour, which could pave the way for potential applications in plasma physics and also in other areas such as transport and conductivity in dense electrolytes \cite{Silkina}. In particular, for dense electrolytes under AC conditions, the effects presented above are typically neglected \cite{Andelman,Rotenberg} and could instead be important to bridge the gap between the current theoretical predictions and the AC conductivity observed at molar concentrations.

All in all, the hyperballistic dynamics under an external AC field discovered here could potentially lead to a new controlled setup for the acceleration of ionized matter \cite{Eubank}.
Future extensions of the current theory will include the effect of magnetic fields for magnetized plasmas and possibly magneto-hydrodynamic effects. 
Also, the present theory may represent the starting point of linear response theories to predict material response properties under external fields \cite{Cui2017,Knezevich,viscosity,zaccone2023theory} or within molecular simulations \cite{Ceriotti}.

\appendix
\section{Average of noise and its autocorrelations}
\label{app:1}
The average force fluctuations on the particle is the sum of all the incoherent contributions on some finite time span $T$, \begin{equation}
\langle F_p(t)\rangle=\frac{1}{T}\int_t^{t+T}F_p(t')dt'.
\label{eq:timeAverage}
\end{equation}
Similarly, the autocorrelation of $F_v$, i.e. the second moment, is expressed as
\begin{equation}
    \langle F_p(t)F_p(t+\tau)\rangle=\frac{1}{T}\int_t^{t+T} F_p(t')F_p(t'+\tau)dt'.
\end{equation}

If the bath is in thermal equilibrium at $t=0$,
since the equilibrium system is ergodic, the average, Eq.~(\ref{eq:timeAverage}), is equivalent to a Boltzmann average on every possible bath configuration $\{x_1,..,x_N,p_1,..,p_N\}$: 
\begin{eqnarray}
\langle F_p(t)\rangle_{H_b^0}&=&\prod_{i=1}^N\int_{-\infty}^{+\infty}dp_ie^{-\frac{p_i^2}{2m_ik_BT}}\times\nonumber\\&&\times\int_{-\infty}^{+\infty}dx_ie^{-\frac{m_i\omega_i^2}{2k_BT}\left(x_i-\frac{\nu_i^2}{\omega_i^2}x\right)^2}F_v(t).\qquad
\label{eq:mediaNulla}
\end{eqnarray}
Completing the squares in the Gaussian integrals, we obtain
\begin{subequations}
    \begin{align}
        \langle p_i\rangle&=0,\\
        \langle p_i^2\rangle&=m_ik_BT,\\
        \Bigg\langle x_i-\frac{\nu_i^2}{\omega_i^2}x\Bigg\rangle&=0\label{eq:mediaPosizioni},\\
        \Bigg\langle\left(x_i-\frac{\nu_i^2}{\omega_i^2}x\right)^2\Bigg\rangle&=\frac{k_BT}{m_i\omega_i^2}.\label{eq:varianzaPosizioni}
    \end{align}
\end{subequations}

When the bath oscillators are subject to the action of the external field, they respond and relax fast (on times $\gamma_0^{-1}\simeq 10^{-13}s$, much shorter than the time-scale $\Omega^{-1}$ on which the transient is observed) relative to the dynamics of the tagged particle, so the oscillatory field can be approximated with its value in a neighborhood of $t=0$~\cite{Zwanzig2001}:
\begin{eqnarray}&&H_b(t=0)\approx \nonumber\\&&~~\ \frac{1}{2}\sum_{i=1}^N\left(\frac{p_i^2}{m_i}+m_i\omega_i^2\left(x_i-\frac{\nu_i^2}{\omega_i^2}x-\frac{q_i}{m_i\omega_i^2}E_0\right)^2\right)~~~\ \end{eqnarray}
plus a constant term that depends on $x$. Since the latter remains inert in the computation of bath averages, it is irrelevant for our results. Then, the average force is
\begin{equation}
\langle F_p(t)\rangle=\prod_{i=1}^N\int_{-\infty}^{+\infty}dx_ie^{-\frac{m_i\omega_i^2}{2k_BT}\left(x_i-\frac{\nu_i^2}{\omega_i^2}x-\frac{q_i}{m_i\omega_i^2}E_0\right)^2}F_v(t).
\end{equation}
Further more, we have
\begin{equation}
\Bigg\langle x_i-\frac{\nu_i^2}{\omega_i^2}x-\frac{q_i}{m_i\omega_i^2}E_0\Bigg\rangle=0
\end{equation}
and 
\begin{equation}
\Bigg\langle\left(x_i-\frac{\nu_i^2}{\omega_i^2}x-\frac{q_i}{m_i\omega_i^2}E_0\right)^2\Bigg\rangle=\frac{k_BT}{m_i\omega_i^2}.
\end{equation} 

Therefore, referring to field-on  (\ref{eq:caseOne}) and field-off (\ref{eq:caseTwo}) ICs, we find the effective field (\ref{eq:campoEffettivo}) is equal to 
\begin{eqnarray}
E_i'(\omega_i,t)&=&\nu_i^2E_0\begin{cases} 
\frac{1}{\omega_i}\frac{\omega_i\sin(\Omega t)-\Omega\sin(\omega_it)}{\omega_i^2-\Omega^2}, &   \text{off,}\ \ \\
\frac{\cos(\Omega t)-\cos(\omega_it)}{\omega_i^2-\Omega^2},&   \text{on.}
\end{cases}
\label{eq:lorentzFieldExplicit}\end{eqnarray}
\begin{widetext}
then for the average force we have
\begin{eqnarray}\langle F_p(t)\rangle&=&\prod_{i=1}^N\int_{-\infty}^{+\infty}dx_iF_p(t)
\begin{cases}
\exp\Bigg[-\frac{m_i\omega_i^2}{2k_BT}\Bigg(x_i-\frac{\nu_i^2}{\omega_i^2}x\Bigg)^2\Bigg],&\text{off,}\ \\
\exp\Bigg[-\frac{m_i\omega_i^2}{2k_BT}\Bigg(x_i-\frac{\nu_i^2}{\omega_i^2}x-\frac{q_iE_0}{m_i\omega_i^2}\Bigg)^2\Bigg],&\text{on.}
\end{cases}\end{eqnarray}
Using formula (\ref{eq:deterministicNoise}) for the noise,
\begin{eqnarray}
\langle F_p(t)\rangle&=&
\sum_{i=1}^N\begin{cases}
q_iE_i'(t),&\text{off,}\\
m_i\nu_i^2\frac{q_iE_0}{m_i\omega_i^2}\cos(\omega_i t)+q_iE_i'(t),&\text{on,}
\end{cases}\nonumber\\&=&
\sum_{i=0}^Nq_i\nu_i^2E_0\begin{cases} 
\frac{1}{\omega_i}\frac{\omega_i\sin(\Omega t)-\Omega\sin(\omega_it)}{\omega_i^2-\Omega^2},&  \text{off,}\\
\frac{\cos(\omega_it)}{\omega_i^2}+
\frac{\cos(\Omega t)-\cos(\omega_it)}{\omega_i^2-\Omega^2},&  \text{on,}
\end{cases}
\nonumber\\&=&
q_0g_0\nu_0^2E_0\int_0^{\omega_D}d\omega\begin{cases} 
\omega \frac{\omega\sin(\Omega t)-\Omega\sin(\omega t)}{\omega^2-\Omega^2},&  \text{off,}\\
\cos(\omega t)+
\omega^2\frac{\cos(\Omega t)-\cos(\omega t)}{\omega^2-\Omega^2},&  \text{on,}
\end{cases}
\end{eqnarray}
\end{widetext}
corresponding to Eqs.~\eqref{eq:dieciCoseni},~\eqref{eq:mediaGeneralizzata2} of the main text. In the small $\Omega$ limit,
\begin{eqnarray}
\langle F_p(t)\rangle&=&\begin{cases} 
      QE_0\sin(\Omega t),&\text{off,}\\
      QE_0\cos(\Omega t),&\text{on,}
\end{cases}\label{eq:A12}
\end{eqnarray} which gives Eqs. (\ref{eq:venti}, ~\ref{eq:ventuno}) of the main text. Moreover, setting $\nu(\omega)=\nu_0$ in  Eq.~\eqref{eq:caricaPolarizzazione}, one finds \begin{eqnarray}
Q=g_0q_0\nu_0^2\omega_D
\end{eqnarray} which is~Eq.~\eqref{eq:caricaVentitre} of the main text.

The autocorrelation function can be similarly evaluated,
leading to Eqs. (\ref{eq:teoremaFluttuazioneZaccone},~\ref{eq:teoremaFluttuazioneGenerale}) of the main text, which for 
$E_0=0$ reduce to Kubo's result~\cite{Kubo1966}:
\begin{equation}\langle F_p(t)F_p(t')\rangle=mk_BTK(t-t').\label{eq:fluctuationDissipation}\end{equation}

\section{Diffusivity}
The mean square displacement of the Brownian particle is
\label{app:2}
\begin{equation}
\langle\{x(t)-x_0\}^2\rangle=\int_0^tdt_1\int_0^tdt_2\langle v(t_1)v(t_2)\rangle.
\label{eq:eqMSD}
\end{equation}

Under field-off ICs, and with a Markovian kernel that obeys the Markovian FDT given in Eq. \eqref{delta}, we get
\begin{align}
\langle\{x(t)-x_0\}^2\rangle=
2\frac{k_BT}{m\gamma_0}t
+\left(\frac{(q+Q)E_0}{\gamma_0 m\Omega}\right)^2(\cos(\Omega t)-1)^2.
\label{eq:MSDEquilibriumCI}
\end{align} The second term is a new contribution due to the polarization of the bath, which may be observable even for large times $t$ if the field's intensity $E_0^2$ is large enough. The diffusivity, $D(t)=\langle\{x(t)-x_0\}^2\rangle/(2t)$, is
\begin{equation}
D(t)=\frac{k_BT}{m\gamma_0}\left(
1+\mathcal{E}_0^2\frac{(\cos(\Omega t)-1)^2}{\Omega t}\right)
\label{eq:diffusivity1-2}
\end{equation}
where we have introduced the dimensionless quantity
\begin{equation}
\mathcal{E}_0=\frac{(q+Q)E_0}{\sqrt{\gamma_0 m\Omega k_BT}}.\label{eq:adimensionalField}
\end{equation}
Eqs. (\ref{eq:diffusivity1-2},~\ref{eq:adimensionalField}) are Eqs. (\ref{eq:diffusivity1},~\ref{eq:E1}) of the main text.\\

Under field-on ICs, and again with the Markovian FDT of Eq. \eqref{delta}, we get
\begin{align}
\langle\{x(t)-x_0\}^2\rangle&=2\frac{k_BT}{m\gamma_0}
t+\left(\frac{(q+Q)E_0}{\gamma_0 m\Omega}\right)^2\sin^2(\Omega t).
\label{eq:MSDNonequilibriumCI}
\end{align}
The diffusivity becomes 
\begin{equation}
D(t)=\frac{k_BT}{m\gamma_0}\left(
1+\mathcal{E}_0^2
\frac{\sin^2(\Omega t)}{\Omega t}
\right),
\label{eq:diffusivity2
}\end{equation}
corresponding to Eq.~\eqref{eq:diffusivity2} of the main text.

\bibliographystyle{apsrev4-1}
\bibliography{refs}

\begin{thebibliography}{57}%
\makeatletter
\providecommand \@ifxundefined [1]{%
 \@ifx{#1\undefined}
}%
\providecommand \@ifnum [1]{%
 \ifnum #1\expandafter \@firstoftwo
 \else \expandafter \@secondoftwo
 \fi
}%
\providecommand \@ifx [1]{%
 \ifx #1\expandafter \@firstoftwo
 \else \expandafter \@secondoftwo
 \fi
}%
\providecommand \natexlab [1]{#1}%
\providecommand \enquote  [1]{``#1''}%
\providecommand \bibnamefont  [1]{#1}%
\providecommand \bibfnamefont [1]{#1}%
\providecommand \citenamefont [1]{#1}%
\providecommand \href@noop [0]{\@secondoftwo}%
\providecommand \href [0]{\begingroup \@sanitize@url \@href}%
\providecommand \@href[1]{\@@startlink{#1}\@@href}%
\providecommand \@@href[1]{\endgroup#1\@@endlink}%
\providecommand \@sanitize@url [0]{\catcode `\\12\catcode `\$12\catcode `\&12\catcode `\#12\catcode `\^12\catcode `\_12\catcode `\%12\relax}%
\providecommand \@@startlink[1]{}%
\providecommand \@@endlink[0]{}%
\providecommand \url  [0]{\begingroup\@sanitize@url \@url }%
\providecommand \@url [1]{\endgroup\@href {#1}{\urlprefix }}%
\providecommand \urlprefix  [0]{URL }%
\providecommand \Eprint [0]{\href }%
\providecommand \doibase [0]{http://dx.doi.org/}%
\providecommand \selectlanguage [0]{\@gobble}%
\providecommand \bibinfo  [0]{\@secondoftwo}%
\providecommand \bibfield  [0]{\@secondoftwo}%
\providecommand \translation [1]{[#1]}%
\providecommand \BibitemOpen [0]{}%
\providecommand \bibitemStop [0]{}%
\providecommand \bibitemNoStop [0]{.\EOS\space}%
\providecommand \EOS [0]{\spacefactor3000\relax}%
\providecommand \BibitemShut  [1]{\csname bibitem#1\endcsname}%
\let\auto@bib@innerbib\@empty
\bibitem [{\citenamefont {Ceriotti}\ \emph {et~al.}(2009)\citenamefont {Ceriotti}, \citenamefont {Bussi},\ and\ \citenamefont {Parrinello}}]{Ceriotti}%
  \BibitemOpen
  \bibfield  {author} {\bibinfo {author} {\bibfnamefont {M.}~\bibnamefont {Ceriotti}}, \bibinfo {author} {\bibfnamefont {G.}~\bibnamefont {Bussi}}, \ and\ \bibinfo {author} {\bibfnamefont {M.}~\bibnamefont {Parrinello}},\ }\href {\doibase 10.1103/PhysRevLett.102.020601} {\bibfield  {journal} {\bibinfo  {journal} {Phys. Rev. Lett.}\ }\textbf {\bibinfo {volume} {102}},\ \bibinfo {pages} {020601} (\bibinfo {year} {2009})}\BibitemShut {NoStop}%
\bibitem [{\citenamefont {Ivlev}\ \emph {et~al.}(2005)\citenamefont {Ivlev}, \citenamefont {Zhdanov}, \citenamefont {Klumov},\ and\ \citenamefont {Morfill}}]{IvZhKl05}%
  \BibitemOpen
  \bibfield  {author} {\bibinfo {author} {\bibfnamefont {A.~V.}\ \bibnamefont {Ivlev}}, \bibinfo {author} {\bibfnamefont {S.~K.}\ \bibnamefont {Zhdanov}}, \bibinfo {author} {\bibfnamefont {B.~A.}\ \bibnamefont {Klumov}}, \ and\ \bibinfo {author} {\bibfnamefont {G.~E.}\ \bibnamefont {Morfill}},\ }\href {\doibase 10.1063/1.2041327} {\bibfield  {journal} {\bibinfo  {journal} {Physics of Plasmas}\ }\textbf {\bibinfo {volume} {12}} (\bibinfo {year} {2005}),\ 10.1063/1.2041327}\BibitemShut {NoStop}%
\bibitem [{\citenamefont {Mabey}\ \emph {et~al.}(2017)\citenamefont {Mabey}, \citenamefont {Richardson}, \citenamefont {White}, \citenamefont {Fletcher}, \citenamefont {Glenzer}, \citenamefont {Hartley}, \citenamefont {Vorberger}, \citenamefont {Gericke},\ and\ \citenamefont {Gregori}}]{Mabey2017}%
  \BibitemOpen
  \bibfield  {author} {\bibinfo {author} {\bibfnamefont {P.}~\bibnamefont {Mabey}}, \bibinfo {author} {\bibfnamefont {S.}~\bibnamefont {Richardson}}, \bibinfo {author} {\bibfnamefont {T.~G.}\ \bibnamefont {White}}, \bibinfo {author} {\bibfnamefont {L.~B.}\ \bibnamefont {Fletcher}}, \bibinfo {author} {\bibfnamefont {S.~H.}\ \bibnamefont {Glenzer}}, \bibinfo {author} {\bibfnamefont {N.~J.}\ \bibnamefont {Hartley}}, \bibinfo {author} {\bibfnamefont {J.}~\bibnamefont {Vorberger}}, \bibinfo {author} {\bibfnamefont {D.~O.}\ \bibnamefont {Gericke}}, \ and\ \bibinfo {author} {\bibfnamefont {G.}~\bibnamefont {Gregori}},\ }\href {\doibase 10.1038/ncomms14125} {\bibfield  {journal} {\bibinfo  {journal} {Nature Communications}\ }\textbf {\bibinfo {volume} {8}},\ \bibinfo {pages} {14125} (\bibinfo {year} {2017})}\BibitemShut {NoStop}%
\bibitem [{\citenamefont {Graziani}\ \emph {et~al.}(2012)\citenamefont {Graziani}, \citenamefont {Batista}, \citenamefont {Benedict}, \citenamefont {Castor}, \citenamefont {Chen}, \citenamefont {Chen}, \citenamefont {Fichtl}, \citenamefont {Glosli}, \citenamefont {Grabowski}, \citenamefont {Graf}, \citenamefont {Hau-Riege}, \citenamefont {Hazi}, \citenamefont {Khairallah}, \citenamefont {Krauss}, \citenamefont {Langdon}, \citenamefont {London}, \citenamefont {Markmann}, \citenamefont {Murillo}, \citenamefont {Richards}, \citenamefont {Scott}, \citenamefont {Shepherd}, \citenamefont {Stanton}, \citenamefont {Streitz}, \citenamefont {Surh}, \citenamefont {Weisheit},\ and\ \citenamefont {Whitley}}]{Graziani}%
  \BibitemOpen
  \bibfield  {author} {\bibinfo {author} {\bibfnamefont {F.~R.}\ \bibnamefont {Graziani}}, \bibinfo {author} {\bibfnamefont {V.~S.}\ \bibnamefont {Batista}}, \bibinfo {author} {\bibfnamefont {L.~X.}\ \bibnamefont {Benedict}}, \bibinfo {author} {\bibfnamefont {J.~I.}\ \bibnamefont {Castor}}, \bibinfo {author} {\bibfnamefont {H.}~\bibnamefont {Chen}}, \bibinfo {author} {\bibfnamefont {S.~N.}\ \bibnamefont {Chen}}, \bibinfo {author} {\bibfnamefont {C.~A.}\ \bibnamefont {Fichtl}}, \bibinfo {author} {\bibfnamefont {J.~N.}\ \bibnamefont {Glosli}}, \bibinfo {author} {\bibfnamefont {P.~E.}\ \bibnamefont {Grabowski}}, \bibinfo {author} {\bibfnamefont {A.~T.}\ \bibnamefont {Graf}}, \bibinfo {author} {\bibfnamefont {S.~P.}\ \bibnamefont {Hau-Riege}}, \bibinfo {author} {\bibfnamefont {A.~U.}\ \bibnamefont {Hazi}}, \bibinfo {author} {\bibfnamefont {S.~A.}\ \bibnamefont {Khairallah}}, \bibinfo {author} {\bibfnamefont {L.}~\bibnamefont {Krauss}}, \bibinfo {author} {\bibfnamefont {A.~B.}\ \bibnamefont {Langdon}}, \bibinfo
  {author} {\bibfnamefont {R.~A.}\ \bibnamefont {London}}, \bibinfo {author} {\bibfnamefont {A.}~\bibnamefont {Markmann}}, \bibinfo {author} {\bibfnamefont {M.~S.}\ \bibnamefont {Murillo}}, \bibinfo {author} {\bibfnamefont {D.~F.}\ \bibnamefont {Richards}}, \bibinfo {author} {\bibfnamefont {H.~A.}\ \bibnamefont {Scott}}, \bibinfo {author} {\bibfnamefont {R.}~\bibnamefont {Shepherd}}, \bibinfo {author} {\bibfnamefont {L.~G.}\ \bibnamefont {Stanton}}, \bibinfo {author} {\bibfnamefont {F.~H.}\ \bibnamefont {Streitz}}, \bibinfo {author} {\bibfnamefont {M.~P.}\ \bibnamefont {Surh}}, \bibinfo {author} {\bibfnamefont {J.~C.}\ \bibnamefont {Weisheit}}, \ and\ \bibinfo {author} {\bibfnamefont {H.~D.}\ \bibnamefont {Whitley}},\ }\href {\doibase https://doi.org/10.1016/j.hedp.2011.06.010} {\bibfield  {journal} {\bibinfo  {journal} {High Energy Density Physics}\ }\textbf {\bibinfo {volume} {8}},\ \bibinfo {pages} {105} (\bibinfo {year} {2012})}\BibitemShut {NoStop}%
\bibitem [{\citenamefont {Melzer}(2019)}]{Me19}%
  \BibitemOpen
  \bibfield  {author} {\bibinfo {author} {\bibfnamefont {A.}~\bibnamefont {Melzer}},\ }\href {\doibase 10.1007/978-3-030-20260-6} {\bibfield  {journal} {\bibinfo  {journal} {Lecture Notes in Physics}\ } (\bibinfo {year} {2019}),\ 10.1007/978-3-030-20260-6}\BibitemShut {NoStop}%
\bibitem [{\citenamefont {Marchenko}\ \emph {et~al.}(2023)\citenamefont {Marchenko}, \citenamefont {Aksenova}, \citenamefont {Marchenko}, \citenamefont {Łuczka},\ and\ \citenamefont {Spiechowicz}}]{MaAkMa23}%
  \BibitemOpen
  \bibfield  {author} {\bibinfo {author} {\bibfnamefont {I.~G.}\ \bibnamefont {Marchenko}}, \bibinfo {author} {\bibfnamefont {V.}~\bibnamefont {Aksenova}}, \bibinfo {author} {\bibfnamefont {I.~I.}\ \bibnamefont {Marchenko}}, \bibinfo {author} {\bibfnamefont {J.}~\bibnamefont {Łuczka}}, \ and\ \bibinfo {author} {\bibfnamefont {J.}~\bibnamefont {Spiechowicz}},\ }\href {\doibase 10.1103/physreve.107.064116} {\bibfield  {journal} {\bibinfo  {journal} {Physical Review E}\ }\textbf {\bibinfo {volume} {107}} (\bibinfo {year} {2023}),\ 10.1103/physreve.107.064116}\BibitemShut {NoStop}%
\bibitem [{\citenamefont {Colla}\ \emph {et~al.}(2018)\citenamefont {Colla}, \citenamefont {Mohanty}, \citenamefont {Nöjd}, \citenamefont {Bialik}, \citenamefont {Riede}, \citenamefont {Schurtenberger},\ and\ \citenamefont {Likos}}]{CoMoNo18}%
  \BibitemOpen
  \bibfield  {author} {\bibinfo {author} {\bibfnamefont {T.}~\bibnamefont {Colla}}, \bibinfo {author} {\bibfnamefont {P.~S.}\ \bibnamefont {Mohanty}}, \bibinfo {author} {\bibfnamefont {S.}~\bibnamefont {Nöjd}}, \bibinfo {author} {\bibfnamefont {E.}~\bibnamefont {Bialik}}, \bibinfo {author} {\bibfnamefont {A.}~\bibnamefont {Riede}}, \bibinfo {author} {\bibfnamefont {P.}~\bibnamefont {Schurtenberger}}, \ and\ \bibinfo {author} {\bibfnamefont {C.~N.}\ \bibnamefont {Likos}},\ }\href {\doibase 10.1021/acsnano.7b08843} {\bibfield  {journal} {\bibinfo  {journal} {ACS Nano}\ }\textbf {\bibinfo {volume} {12}},\ \bibinfo {pages} {4321–4337} (\bibinfo {year} {2018})}\BibitemShut {NoStop}%
\bibitem [{\citenamefont {Katzmeier}\ \emph {et~al.}(2022)\citenamefont {Katzmeier}, \citenamefont {Altaner}, \citenamefont {List}, \citenamefont {Gerland},\ and\ \citenamefont {Simmel}}]{KaAlLi22}%
  \BibitemOpen
  \bibfield  {author} {\bibinfo {author} {\bibfnamefont {F.}~\bibnamefont {Katzmeier}}, \bibinfo {author} {\bibfnamefont {B.}~\bibnamefont {Altaner}}, \bibinfo {author} {\bibfnamefont {J.}~\bibnamefont {List}}, \bibinfo {author} {\bibfnamefont {U.}~\bibnamefont {Gerland}}, \ and\ \bibinfo {author} {\bibfnamefont {F.~C.}\ \bibnamefont {Simmel}},\ }\href {\doibase 10.1103/physrevlett.128.058002} {\bibfield  {journal} {\bibinfo  {journal} {Physical Review Letters}\ }\textbf {\bibinfo {volume} {128}} (\bibinfo {year} {2022}),\ 10.1103/physrevlett.128.058002}\BibitemShut {NoStop}%
\bibitem [{\citenamefont {Hoang Ngoc~Minh}\ \emph {et~al.}(2023)\citenamefont {Hoang Ngoc~Minh}, \citenamefont {Stoltz},\ and\ \citenamefont {Rotenberg}}]{Rotenberg}%
  \BibitemOpen
  \bibfield  {author} {\bibinfo {author} {\bibfnamefont {T.}~\bibnamefont {Hoang Ngoc~Minh}}, \bibinfo {author} {\bibfnamefont {G.}~\bibnamefont {Stoltz}}, \ and\ \bibinfo {author} {\bibfnamefont {B.}~\bibnamefont {Rotenberg}},\ }\href {\doibase 10.1063/5.0139258} {\bibfield  {journal} {\bibinfo  {journal} {The Journal of Chemical Physics}\ }\textbf {\bibinfo {volume} {158}},\ \bibinfo {pages} {104103} (\bibinfo {year} {2023})},\ \Eprint {http://arxiv.org/abs/https://pubs.aip.org/aip/jcp/article-pdf/doi/10.1063/5.0139258/16790539/104103\_1\_online.pdf} {https://pubs.aip.org/aip/jcp/article-pdf/doi/10.1063/5.0139258/16790539/104103\_1\_online.pdf} \BibitemShut {NoStop}%
\bibitem [{\citenamefont {Vater}\ \emph {et~al.}(2022)\citenamefont {Vater}, \citenamefont {Isele}, \citenamefont {Siems},\ and\ \citenamefont {Nielaba}}]{VaIsSi22}%
  \BibitemOpen
  \bibfield  {author} {\bibinfo {author} {\bibfnamefont {T.}~\bibnamefont {Vater}}, \bibinfo {author} {\bibfnamefont {M.}~\bibnamefont {Isele}}, \bibinfo {author} {\bibfnamefont {U.}~\bibnamefont {Siems}}, \ and\ \bibinfo {author} {\bibfnamefont {P.}~\bibnamefont {Nielaba}},\ }\href {\doibase 10.1103/physreve.106.024606} {\bibfield  {journal} {\bibinfo  {journal} {Physical Review E}\ }\textbf {\bibinfo {volume} {106}} (\bibinfo {year} {2022}),\ 10.1103/physreve.106.024606}\BibitemShut {NoStop}%
\bibitem [{\citenamefont {Vissers}\ \emph {et~al.}(2011)\citenamefont {Vissers}, \citenamefont {van Blaaderen},\ and\ \citenamefont {Imhof}}]{VivaIm11}%
  \BibitemOpen
  \bibfield  {author} {\bibinfo {author} {\bibfnamefont {T.}~\bibnamefont {Vissers}}, \bibinfo {author} {\bibfnamefont {A.}~\bibnamefont {van Blaaderen}}, \ and\ \bibinfo {author} {\bibfnamefont {A.}~\bibnamefont {Imhof}},\ }\href {\doibase 10.1103/physrevlett.106.228303} {\bibfield  {journal} {\bibinfo  {journal} {Physical Review Letters}\ }\textbf {\bibinfo {volume} {106}} (\bibinfo {year} {2011}),\ 10.1103/physrevlett.106.228303}\BibitemShut {NoStop}%
\bibitem [{\citenamefont {Sütterlin}\ \emph {et~al.}(2009)\citenamefont {Sütterlin}, \citenamefont {Wysocki}, \citenamefont {Ivlev}, \citenamefont {Räth}, \citenamefont {Thomas}, \citenamefont {Rubin-Zuzic}, \citenamefont {Goedheer}, \citenamefont {Fortov}, \citenamefont {Lipaev}, \citenamefont {Molotkov}, \citenamefont {Petrov}, \citenamefont {Morfill},\ and\ \citenamefont {Löwen}}]{SuWyIv09}%
  \BibitemOpen
  \bibfield  {author} {\bibinfo {author} {\bibfnamefont {K.~R.}\ \bibnamefont {Sütterlin}}, \bibinfo {author} {\bibfnamefont {A.}~\bibnamefont {Wysocki}}, \bibinfo {author} {\bibfnamefont {A.~V.}\ \bibnamefont {Ivlev}}, \bibinfo {author} {\bibfnamefont {C.}~\bibnamefont {Räth}}, \bibinfo {author} {\bibfnamefont {H.~M.}\ \bibnamefont {Thomas}}, \bibinfo {author} {\bibfnamefont {M.}~\bibnamefont {Rubin-Zuzic}}, \bibinfo {author} {\bibfnamefont {W.~J.}\ \bibnamefont {Goedheer}}, \bibinfo {author} {\bibfnamefont {V.~E.}\ \bibnamefont {Fortov}}, \bibinfo {author} {\bibfnamefont {A.~M.}\ \bibnamefont {Lipaev}}, \bibinfo {author} {\bibfnamefont {V.~I.}\ \bibnamefont {Molotkov}}, \bibinfo {author} {\bibfnamefont {O.~F.}\ \bibnamefont {Petrov}}, \bibinfo {author} {\bibfnamefont {G.~E.}\ \bibnamefont {Morfill}}, \ and\ \bibinfo {author} {\bibfnamefont {H.}~\bibnamefont {Löwen}},\ }\href {\doibase 10.1103/physrevlett.102.085003} {\bibfield  {journal} {\bibinfo  {journal} {Physical Review Letters}\ }\textbf {\bibinfo
  {volume} {102}} (\bibinfo {year} {2009}),\ 10.1103/physrevlett.102.085003}\BibitemShut {NoStop}%
\bibitem [{\citenamefont {Huang}\ \emph {et~al.}(2022)\citenamefont {Huang}, \citenamefont {Nosenko}, \citenamefont {Huang-Fu}, \citenamefont {Thomas},\ and\ \citenamefont {Du}}]{HuNoHu22}%
  \BibitemOpen
  \bibfield  {author} {\bibinfo {author} {\bibfnamefont {H.}~\bibnamefont {Huang}}, \bibinfo {author} {\bibfnamefont {V.}~\bibnamefont {Nosenko}}, \bibinfo {author} {\bibfnamefont {H.-X.}\ \bibnamefont {Huang-Fu}}, \bibinfo {author} {\bibfnamefont {H.~M.}\ \bibnamefont {Thomas}}, \ and\ \bibinfo {author} {\bibfnamefont {C.-R.}\ \bibnamefont {Du}},\ }\href {\doibase 10.1063/5.0096938} {\bibfield  {journal} {\bibinfo  {journal} {Physics of Plasmas}\ }\textbf {\bibinfo {volume} {29}} (\bibinfo {year} {2022}),\ 10.1063/5.0096938}\BibitemShut {NoStop}%
\bibitem [{\citenamefont {Hansen}\ and\ \citenamefont {Sj\"{o}gren}(1982)}]{Hansen1982}%
  \BibitemOpen
  \bibfield  {author} {\bibinfo {author} {\bibfnamefont {J.~P.}\ \bibnamefont {Hansen}}\ and\ \bibinfo {author} {\bibfnamefont {L.}~\bibnamefont {Sj\"{o}gren}},\ }\href {\doibase 10.1063/1.863808} {\bibfield  {journal} {\bibinfo  {journal} {The Physics of Fluids}\ }\textbf {\bibinfo {volume} {25}},\ \bibinfo {pages} {617–628} (\bibinfo {year} {1982})}\BibitemShut {NoStop}%
\bibitem [{\citenamefont {Hagstrom}\ and\ \citenamefont {Morrison}(2011)}]{Hagstrom2011}%
  \BibitemOpen
  \bibfield  {author} {\bibinfo {author} {\bibfnamefont {G.~I.}\ \bibnamefont {Hagstrom}}\ and\ \bibinfo {author} {\bibfnamefont {P.}~\bibnamefont {Morrison}},\ }\href {\doibase 10.1016/j.physd.2011.02.007} {\bibfield  {journal} {\bibinfo  {journal} {Physica D: Nonlinear Phenomena}\ }\textbf {\bibinfo {volume} {240}},\ \bibinfo {pages} {1652–1660} (\bibinfo {year} {2011})}\BibitemShut {NoStop}%
\bibitem [{\citenamefont {Dan}\ \emph {et~al.}(2017)\citenamefont {Dan}, \citenamefont {Guo}, \citenamefont {Li}, \citenamefont {Chen}, \citenamefont {Yan},\ and\ \citenamefont {Huang}}]{Dan2017}%
  \BibitemOpen
  \bibfield  {author} {\bibinfo {author} {\bibfnamefont {L.}~\bibnamefont {Dan}}, \bibinfo {author} {\bibfnamefont {L.-X.}\ \bibnamefont {Guo}}, \bibinfo {author} {\bibfnamefont {J.-T.}\ \bibnamefont {Li}}, \bibinfo {author} {\bibfnamefont {W.}~\bibnamefont {Chen}}, \bibinfo {author} {\bibfnamefont {X.}~\bibnamefont {Yan}}, \ and\ \bibinfo {author} {\bibfnamefont {Q.-Q.}\ \bibnamefont {Huang}},\ }\href {\doibase 10.1063/1.4999499} {\bibfield  {journal} {\bibinfo  {journal} {Physics of Plasmas}\ }\textbf {\bibinfo {volume} {24}} (\bibinfo {year} {2017}),\ 10.1063/1.4999499}\BibitemShut {NoStop}%
\bibitem [{\citenamefont {Jia}\ \emph {et~al.}(2016)\citenamefont {Jia}, \citenamefont {Yuan}, \citenamefont {Liu}, \citenamefont {Yue}, \citenamefont {Gao}, \citenamefont {Wang}, \citenamefont {Zhou}, \citenamefont {Wu},\ and\ \citenamefont {Li}}]{Jia2016}%
  \BibitemOpen
  \bibfield  {author} {\bibinfo {author} {\bibfnamefont {J.}~\bibnamefont {Jia}}, \bibinfo {author} {\bibfnamefont {C.}~\bibnamefont {Yuan}}, \bibinfo {author} {\bibfnamefont {S.}~\bibnamefont {Liu}}, \bibinfo {author} {\bibfnamefont {F.}~\bibnamefont {Yue}}, \bibinfo {author} {\bibfnamefont {R.}~\bibnamefont {Gao}}, \bibinfo {author} {\bibfnamefont {Y.}~\bibnamefont {Wang}}, \bibinfo {author} {\bibfnamefont {Z.-X.}\ \bibnamefont {Zhou}}, \bibinfo {author} {\bibfnamefont {J.}~\bibnamefont {Wu}}, \ and\ \bibinfo {author} {\bibfnamefont {H.}~\bibnamefont {Li}},\ }\href {\doibase 10.1063/1.4946780} {\bibfield  {journal} {\bibinfo  {journal} {Physics of Plasmas}\ }\textbf {\bibinfo {volume} {23}} (\bibinfo {year} {2016}),\ 10.1063/1.4946780}\BibitemShut {NoStop}%
\bibitem [{\citenamefont {Cui}\ and\ \citenamefont {Zaccone}(2018)}]{CuiZaccone2018}%
  \BibitemOpen
  \bibfield  {author} {\bibinfo {author} {\bibfnamefont {B.}~\bibnamefont {Cui}}\ and\ \bibinfo {author} {\bibfnamefont {A.}~\bibnamefont {Zaccone}},\ }\href {\doibase 10.1103/PhysRevE.97.060102} {\bibfield  {journal} {\bibinfo  {journal} {Phys. Rev. E}\ }\textbf {\bibinfo {volume} {97}},\ \bibinfo {pages} {060102} (\bibinfo {year} {2018})}\BibitemShut {NoStop}%
\bibitem [{\citenamefont {Grabert}\ \emph {et~al.}(2016)\citenamefont {Grabert}, \citenamefont {Nalbach}, \citenamefont {Reichert},\ and\ \citenamefont {Thorwart}}]{Thorwart_2016}%
  \BibitemOpen
  \bibfield  {author} {\bibinfo {author} {\bibfnamefont {H.}~\bibnamefont {Grabert}}, \bibinfo {author} {\bibfnamefont {P.}~\bibnamefont {Nalbach}}, \bibinfo {author} {\bibfnamefont {J.}~\bibnamefont {Reichert}}, \ and\ \bibinfo {author} {\bibfnamefont {M.}~\bibnamefont {Thorwart}},\ }\href {\doibase 10.1021/acs.jpclett.6b00703} {\bibfield  {journal} {\bibinfo  {journal} {The Journal of Physical Chemistry Letters}\ }\textbf {\bibinfo {volume} {7}},\ \bibinfo {pages} {2015} (\bibinfo {year} {2016})},\ \bibinfo {note} {pMID: 27176818},\ \Eprint {http://arxiv.org/abs/https://doi.org/10.1021/acs.jpclett.6b00703} {https://doi.org/10.1021/acs.jpclett.6b00703} \BibitemShut {NoStop}%
\bibitem [{\citenamefont {Grabert}\ and\ \citenamefont {Thorwart}(2018)}]{Thorwart_2018}%
  \BibitemOpen
  \bibfield  {author} {\bibinfo {author} {\bibfnamefont {H.}~\bibnamefont {Grabert}}\ and\ \bibinfo {author} {\bibfnamefont {M.}~\bibnamefont {Thorwart}},\ }\href {\doibase 10.1103/PhysRevE.98.012122} {\bibfield  {journal} {\bibinfo  {journal} {Phys. Rev. E}\ }\textbf {\bibinfo {volume} {98}},\ \bibinfo {pages} {012122} (\bibinfo {year} {2018})}\BibitemShut {NoStop}%
\bibitem [{\citenamefont {Barnaveli}\ and\ \citenamefont {van Roij}(2024{\natexlab{a}})}]{vanRoij}%
  \BibitemOpen
  \bibfield  {author} {\bibinfo {author} {\bibfnamefont {A.}~\bibnamefont {Barnaveli}}\ and\ \bibinfo {author} {\bibfnamefont {R.}~\bibnamefont {van Roij}},\ }\href {\doibase 10.1021/acs.langmuir.4c01516} {\bibfield  {journal} {\bibinfo  {journal} {Langmuir}\ }\textbf {\bibinfo {volume} {40}},\ \bibinfo {pages} {14066} (\bibinfo {year} {2024}{\natexlab{a}})},\ \bibinfo {note} {pMID: 38916199},\ \Eprint {http://arxiv.org/abs/https://doi.org/10.1021/acs.langmuir.4c01516} {https://doi.org/10.1021/acs.langmuir.4c01516} \BibitemShut {NoStop}%
\bibitem [{\citenamefont {Barnaveli}\ and\ \citenamefont {van Roij}(2024{\natexlab{b}})}]{vanRoij2}%
  \BibitemOpen
  \bibfield  {author} {\bibinfo {author} {\bibfnamefont {A.}~\bibnamefont {Barnaveli}}\ and\ \bibinfo {author} {\bibfnamefont {R.}~\bibnamefont {van Roij}},\ }\href {\doibase 10.1039/D3SM01306E} {\bibfield  {journal} {\bibinfo  {journal} {Soft Matter}\ }\textbf {\bibinfo {volume} {20}},\ \bibinfo {pages} {704} (\bibinfo {year} {2024}{\natexlab{b}})}\BibitemShut {NoStop}%
\bibitem [{\citenamefont {Pireddu}\ and\ \citenamefont {Rotenberg}(2023)}]{PhysRevLett.130.098001}%
  \BibitemOpen
  \bibfield  {author} {\bibinfo {author} {\bibfnamefont {G.}~\bibnamefont {Pireddu}}\ and\ \bibinfo {author} {\bibfnamefont {B.}~\bibnamefont {Rotenberg}},\ }\href {\doibase 10.1103/PhysRevLett.130.098001} {\bibfield  {journal} {\bibinfo  {journal} {Phys. Rev. Lett.}\ }\textbf {\bibinfo {volume} {130}},\ \bibinfo {pages} {098001} (\bibinfo {year} {2023})}\BibitemShut {NoStop}%
\bibitem [{\citenamefont {Zwanzig}(2001)}]{Zwanzig2001}%
  \BibitemOpen
  \bibfield  {author} {\bibinfo {author} {\bibfnamefont {R.~W.}\ \bibnamefont {Zwanzig}},\ }\href@noop {} {\emph {\bibinfo {title} {Nonequilibrium Statistical Mechanics}}}\ (\bibinfo  {publisher} {Oxford University Press},\ \bibinfo {address} {New York, NY},\ \bibinfo {year} {2001})\BibitemShut {NoStop}%
\bibitem [{\citenamefont {Vaulina}\ \emph {et~al.}(1999)\citenamefont {Vaulina}, \citenamefont {Khrapak}, \citenamefont {Nefedov},\ and\ \citenamefont {Petrov}}]{Vaulina1999}%
  \BibitemOpen
  \bibfield  {author} {\bibinfo {author} {\bibfnamefont {O.~S.}\ \bibnamefont {Vaulina}}, \bibinfo {author} {\bibfnamefont {S.~A.}\ \bibnamefont {Khrapak}}, \bibinfo {author} {\bibfnamefont {A.~P.}\ \bibnamefont {Nefedov}}, \ and\ \bibinfo {author} {\bibfnamefont {O.~F.}\ \bibnamefont {Petrov}},\ }\href {\doibase 10.1103/PhysRevE.60.5959} {\bibfield  {journal} {\bibinfo  {journal} {Phys. Rev. E}\ }\textbf {\bibinfo {volume} {60}},\ \bibinfo {pages} {5959} (\bibinfo {year} {1999})}\BibitemShut {NoStop}%
\bibitem [{\citenamefont {Ivlev}\ \emph {et~al.}(2000)\citenamefont {Ivlev}, \citenamefont {Konopka},\ and\ \citenamefont {Morfill}}]{Ivlev2000}%
  \BibitemOpen
  \bibfield  {author} {\bibinfo {author} {\bibfnamefont {A.~V.}\ \bibnamefont {Ivlev}}, \bibinfo {author} {\bibfnamefont {U.}~\bibnamefont {Konopka}}, \ and\ \bibinfo {author} {\bibfnamefont {G.}~\bibnamefont {Morfill}},\ }\href {\doibase 10.1103/PhysRevE.62.2739} {\bibfield  {journal} {\bibinfo  {journal} {Phys. Rev. E}\ }\textbf {\bibinfo {volume} {62}},\ \bibinfo {pages} {2739} (\bibinfo {year} {2000})}\BibitemShut {NoStop}%
\bibitem [{\citenamefont {Khrapak}\ and\ \citenamefont {Morfill}(2002)}]{Morfill2002}%
  \BibitemOpen
  \bibfield  {author} {\bibinfo {author} {\bibfnamefont {S.~A.}\ \bibnamefont {Khrapak}}\ and\ \bibinfo {author} {\bibfnamefont {G.~E.}\ \bibnamefont {Morfill}},\ }\href {\doibase 10.1063/1.1431248} {\bibfield  {journal} {\bibinfo  {journal} {Physics of Plasmas}\ }\textbf {\bibinfo {volume} {9}},\ \bibinfo {pages} {619} (\bibinfo {year} {2002})},\ \Eprint {http://arxiv.org/abs/https://pubs.aip.org/aip/pop/article-pdf/9/2/619/19189224/619\_1\_online.pdf} {https://pubs.aip.org/aip/pop/article-pdf/9/2/619/19189224/619\_1\_online.pdf} \BibitemShut {NoStop}%
\bibitem [{\citenamefont {Caldeira}\ and\ \citenamefont {Leggett}(1983)}]{Caldeira1983}%
  \BibitemOpen
  \bibfield  {author} {\bibinfo {author} {\bibfnamefont {A.}~\bibnamefont {Caldeira}}\ and\ \bibinfo {author} {\bibfnamefont {A.}~\bibnamefont {Leggett}},\ }\href {\doibase https://doi.org/10.1016/0003-4916(83)90202-6} {\bibfield  {journal} {\bibinfo  {journal} {Annals of Physics}\ }\textbf {\bibinfo {volume} {149}},\ \bibinfo {pages} {374} (\bibinfo {year} {1983})}\BibitemShut {NoStop}%
\bibitem [{\citenamefont {Khrapak}(2024)}]{Khrapak}%
  \BibitemOpen
  \bibfield  {author} {\bibinfo {author} {\bibfnamefont {S.}~\bibnamefont {Khrapak}},\ }\href {\doibase https://doi.org/10.1016/j.physrep.2023.11.004} {\bibfield  {journal} {\bibinfo  {journal} {Physics Reports}\ }\textbf {\bibinfo {volume} {1050}},\ \bibinfo {pages} {1} (\bibinfo {year} {2024})},\ \bibinfo {note} {elementary vibrational model for transport properties of dense fluids}\BibitemShut {NoStop}%
\bibitem [{\citenamefont {Jackson}(1967)}]{Jackson1998}%
  \BibitemOpen
  \bibfield  {author} {\bibinfo {author} {\bibfnamefont {J.~D.}\ \bibnamefont {Jackson}},\ }\href@noop {} {\emph {\bibinfo {title} {Classical Electrodynamics}}},\ \bibinfo {edition} {1st}\ ed.\ (\bibinfo  {publisher} {John Wiley \& Sons},\ \bibinfo {address} {New York, NY},\ \bibinfo {year} {1967})\BibitemShut {NoStop}%
\bibitem [{\citenamefont {Di~Terlizzi}\ \emph {et~al.}(2020)\citenamefont {Di~Terlizzi}, \citenamefont {Ritort},\ and\ \citenamefont {Baiesi}}]{Ritort2020}%
  \BibitemOpen
  \bibfield  {author} {\bibinfo {author} {\bibfnamefont {I.}~\bibnamefont {Di~Terlizzi}}, \bibinfo {author} {\bibfnamefont {F.}~\bibnamefont {Ritort}}, \ and\ \bibinfo {author} {\bibfnamefont {M.}~\bibnamefont {Baiesi}},\ }\href@noop {} {\bibfield  {journal} {\bibinfo  {journal} {J. Stat. Phys.}\ }\textbf {\bibinfo {volume} {181}},\ \bibinfo {pages} {1609} (\bibinfo {year} {2020})}\BibitemShut {NoStop}%
\bibitem [{\citenamefont {Robson}\ \emph {et~al.}(1997)\citenamefont {Robson}, \citenamefont {White},\ and\ \citenamefont {Makabe}}]{Makabe1997}%
  \BibitemOpen
  \bibfield  {author} {\bibinfo {author} {\bibfnamefont {R.}~\bibnamefont {Robson}}, \bibinfo {author} {\bibfnamefont {R.}~\bibnamefont {White}}, \ and\ \bibinfo {author} {\bibfnamefont {T.}~\bibnamefont {Makabe}},\ }\href {\doibase https://doi.org/10.1006/aphy.1997.5733} {\bibfield  {journal} {\bibinfo  {journal} {Annals of Physics}\ }\textbf {\bibinfo {volume} {261}},\ \bibinfo {pages} {74} (\bibinfo {year} {1997})}\BibitemShut {NoStop}%
\bibitem [{\citenamefont {Probstein}(1989)}]{Probstein1989}%
  \BibitemOpen
  \bibfield  {author} {\bibinfo {author} {\bibfnamefont {R.~F.}\ \bibnamefont {Probstein}},\ }\href@noop {} {\emph {\bibinfo {title} {Physicochemical Hydrodynamics}}}\ (\bibinfo  {publisher} {Elsevier},\ \bibinfo {year} {1989})\BibitemShut {NoStop}%
\bibitem [{\citenamefont {Brünig}\ \emph {et~al.}(2022)\citenamefont {Brünig}, \citenamefont {Daldrop},\ and\ \citenamefont {Netz}}]{Florian}%
  \BibitemOpen
  \bibfield  {author} {\bibinfo {author} {\bibfnamefont {F.~N.}\ \bibnamefont {Brünig}}, \bibinfo {author} {\bibfnamefont {J.~O.}\ \bibnamefont {Daldrop}}, \ and\ \bibinfo {author} {\bibfnamefont {R.~R.}\ \bibnamefont {Netz}},\ }\href {\doibase 10.1021/acs.jpcb.2c05923} {\bibfield  {journal} {\bibinfo  {journal} {The Journal of Physical Chemistry B}\ }\textbf {\bibinfo {volume} {126}},\ \bibinfo {pages} {10295} (\bibinfo {year} {2022})},\ \bibinfo {note} {pMID: 36473702},\ \Eprint {http://arxiv.org/abs/https://doi.org/10.1021/acs.jpcb.2c05923} {https://doi.org/10.1021/acs.jpcb.2c05923} \BibitemShut {NoStop}%
\bibitem [{\citenamefont {Bisai}\ and\ \citenamefont {Sen}(2023)}]{Bisai2023}%
  \BibitemOpen
  \bibfield  {author} {\bibinfo {author} {\bibfnamefont {N.}~\bibnamefont {Bisai}}\ and\ \bibinfo {author} {\bibfnamefont {A.}~\bibnamefont {Sen}},\ }\href {\doibase 10.1007/s41614-023-00124-5} {\bibfield  {journal} {\bibinfo  {journal} {Reviews of Modern Plasma Physics}\ }\textbf {\bibinfo {volume} {7}},\ \bibinfo {pages} {22} (\bibinfo {year} {2023})}\BibitemShut {NoStop}%
\bibitem [{\citenamefont {Furno}\ \emph {et~al.}(2008)\citenamefont {Furno}, \citenamefont {Labit}, \citenamefont {Podest\`a}, \citenamefont {Fasoli}, \citenamefont {M\"uller}, \citenamefont {Poli}, \citenamefont {Ricci}, \citenamefont {Theiler}, \citenamefont {Brunner}, \citenamefont {Diallo},\ and\ \citenamefont {Graves}}]{TheilerPRL}%
  \BibitemOpen
  \bibfield  {author} {\bibinfo {author} {\bibfnamefont {I.}~\bibnamefont {Furno}}, \bibinfo {author} {\bibfnamefont {B.}~\bibnamefont {Labit}}, \bibinfo {author} {\bibfnamefont {M.}~\bibnamefont {Podest\`a}}, \bibinfo {author} {\bibfnamefont {A.}~\bibnamefont {Fasoli}}, \bibinfo {author} {\bibfnamefont {S.~H.}\ \bibnamefont {M\"uller}}, \bibinfo {author} {\bibfnamefont {F.~M.}\ \bibnamefont {Poli}}, \bibinfo {author} {\bibfnamefont {P.}~\bibnamefont {Ricci}}, \bibinfo {author} {\bibfnamefont {C.}~\bibnamefont {Theiler}}, \bibinfo {author} {\bibfnamefont {S.}~\bibnamefont {Brunner}}, \bibinfo {author} {\bibfnamefont {A.}~\bibnamefont {Diallo}}, \ and\ \bibinfo {author} {\bibfnamefont {J.}~\bibnamefont {Graves}},\ }\href {\doibase 10.1103/PhysRevLett.100.055004} {\bibfield  {journal} {\bibinfo  {journal} {Phys. Rev. Lett.}\ }\textbf {\bibinfo {volume} {100}},\ \bibinfo {pages} {055004} (\bibinfo {year} {2008})}\BibitemShut {NoStop}%
\bibitem [{\citenamefont {Ricci}\ \emph {et~al.}(2012)\citenamefont {Ricci}, \citenamefont {Halpern}, \citenamefont {Jolliet}, \citenamefont {Loizu}, \citenamefont {Mosetto}, \citenamefont {Fasoli}, \citenamefont {Furno},\ and\ \citenamefont {Theiler}}]{Ricci_2012}%
  \BibitemOpen
  \bibfield  {author} {\bibinfo {author} {\bibfnamefont {P.}~\bibnamefont {Ricci}}, \bibinfo {author} {\bibfnamefont {F.~D.}\ \bibnamefont {Halpern}}, \bibinfo {author} {\bibfnamefont {S.}~\bibnamefont {Jolliet}}, \bibinfo {author} {\bibfnamefont {J.}~\bibnamefont {Loizu}}, \bibinfo {author} {\bibfnamefont {A.}~\bibnamefont {Mosetto}}, \bibinfo {author} {\bibfnamefont {A.}~\bibnamefont {Fasoli}}, \bibinfo {author} {\bibfnamefont {I.}~\bibnamefont {Furno}}, \ and\ \bibinfo {author} {\bibfnamefont {C.}~\bibnamefont {Theiler}},\ }\href {\doibase 10.1088/0741-3335/54/12/124047} {\bibfield  {journal} {\bibinfo  {journal} {Plasma Physics and Controlled Fusion}\ }\textbf {\bibinfo {volume} {54}},\ \bibinfo {pages} {124047} (\bibinfo {year} {2012})}\BibitemShut {NoStop}%
\bibitem [{\citenamefont {Metzler}\ \emph {et~al.}(2014)\citenamefont {Metzler}, \citenamefont {Jeon}, \citenamefont {Cherstvy},\ and\ \citenamefont {Barkai}}]{Metzler}%
  \BibitemOpen
  \bibfield  {author} {\bibinfo {author} {\bibfnamefont {R.}~\bibnamefont {Metzler}}, \bibinfo {author} {\bibfnamefont {J.-H.}\ \bibnamefont {Jeon}}, \bibinfo {author} {\bibfnamefont {A.~G.}\ \bibnamefont {Cherstvy}}, \ and\ \bibinfo {author} {\bibfnamefont {E.}~\bibnamefont {Barkai}},\ }\href {\doibase 10.1039/C4CP03465A} {\bibfield  {journal} {\bibinfo  {journal} {Phys. Chem. Chem. Phys.}\ }\textbf {\bibinfo {volume} {16}},\ \bibinfo {pages} {24128} (\bibinfo {year} {2014})}\BibitemShut {NoStop}%
\bibitem [{\citenamefont {Bouchaud}\ and\ \citenamefont {Georges}(1990)}]{Bouchaud1990}%
  \BibitemOpen
  \bibfield  {author} {\bibinfo {author} {\bibfnamefont {J.-P.}\ \bibnamefont {Bouchaud}}\ and\ \bibinfo {author} {\bibfnamefont {A.}~\bibnamefont {Georges}},\ }\href {\doibase 10.1016/0370-1573(90)90099-n} {\bibfield  {journal} {\bibinfo  {journal} {Physics Reports}\ }\textbf {\bibinfo {volume} {195}},\ \bibinfo {pages} {127–293} (\bibinfo {year} {1990})}\BibitemShut {NoStop}%
\bibitem [{\citenamefont {Ilievski}\ \emph {et~al.}(2021)\citenamefont {Ilievski}, \citenamefont {De~Nardis}, \citenamefont {Gopalakrishnan}, \citenamefont {Vasseur},\ and\ \citenamefont {Ware}}]{superuniv}%
  \BibitemOpen
  \bibfield  {author} {\bibinfo {author} {\bibfnamefont {E.}~\bibnamefont {Ilievski}}, \bibinfo {author} {\bibfnamefont {J.}~\bibnamefont {De~Nardis}}, \bibinfo {author} {\bibfnamefont {S.}~\bibnamefont {Gopalakrishnan}}, \bibinfo {author} {\bibfnamefont {R.}~\bibnamefont {Vasseur}}, \ and\ \bibinfo {author} {\bibfnamefont {B.}~\bibnamefont {Ware}},\ }\href {\doibase 10.1103/PhysRevX.11.031023} {\bibfield  {journal} {\bibinfo  {journal} {Phys. Rev. X}\ }\textbf {\bibinfo {volume} {11}},\ \bibinfo {pages} {031023} (\bibinfo {year} {2021})}\BibitemShut {NoStop}%
\bibitem [{\citenamefont {Siegle}\ \emph {et~al.}(2010)\citenamefont {Siegle}, \citenamefont {Goychuk},\ and\ \citenamefont {H\"anggi}}]{Haenggi}%
  \BibitemOpen
  \bibfield  {author} {\bibinfo {author} {\bibfnamefont {P.}~\bibnamefont {Siegle}}, \bibinfo {author} {\bibfnamefont {I.}~\bibnamefont {Goychuk}}, \ and\ \bibinfo {author} {\bibfnamefont {P.}~\bibnamefont {H\"anggi}},\ }\href {\doibase 10.1103/PhysRevLett.105.100602} {\bibfield  {journal} {\bibinfo  {journal} {Phys. Rev. Lett.}\ }\textbf {\bibinfo {volume} {105}},\ \bibinfo {pages} {100602} (\bibinfo {year} {2010})}\BibitemShut {NoStop}%
\bibitem [{\citenamefont {Flekkøy}\ \emph {et~al.}(2021)\citenamefont {Flekkøy}, \citenamefont {Hansen},\ and\ \citenamefont {Baldelli}}]{Baldelli}%
  \BibitemOpen
  \bibfield  {author} {\bibinfo {author} {\bibfnamefont {E.~G.}\ \bibnamefont {Flekkøy}}, \bibinfo {author} {\bibfnamefont {A.}~\bibnamefont {Hansen}}, \ and\ \bibinfo {author} {\bibfnamefont {B.}~\bibnamefont {Baldelli}},\ }\href {\doibase 10.3389/fphy.2021.640560} {\bibfield  {journal} {\bibinfo  {journal} {Frontiers in Physics}\ }\textbf {\bibinfo {volume} {9}} (\bibinfo {year} {2021}),\ 10.3389/fphy.2021.640560}\BibitemShut {NoStop}%
\bibitem [{\citenamefont {Levi}\ \emph {et~al.}(2012)\citenamefont {Levi}, \citenamefont {Krivolapov}, \citenamefont {Fishman},\ and\ \citenamefont {Segev}}]{Levi2012}%
  \BibitemOpen
  \bibfield  {author} {\bibinfo {author} {\bibfnamefont {L.}~\bibnamefont {Levi}}, \bibinfo {author} {\bibfnamefont {Y.}~\bibnamefont {Krivolapov}}, \bibinfo {author} {\bibfnamefont {S.}~\bibnamefont {Fishman}}, \ and\ \bibinfo {author} {\bibfnamefont {M.}~\bibnamefont {Segev}},\ }\href {\doibase 10.1038/nphys2463} {\bibfield  {journal} {\bibinfo  {journal} {Nature Physics}\ }\textbf {\bibinfo {volume} {8}},\ \bibinfo {pages} {912} (\bibinfo {year} {2012})}\BibitemShut {NoStop}%
\bibitem [{\citenamefont {Peccianti}\ and\ \citenamefont {Morandotti}(2012)}]{Peccianti2012}%
  \BibitemOpen
  \bibfield  {author} {\bibinfo {author} {\bibfnamefont {M.}~\bibnamefont {Peccianti}}\ and\ \bibinfo {author} {\bibfnamefont {R.}~\bibnamefont {Morandotti}},\ }\href {\doibase 10.1038/nphys2486} {\bibfield  {journal} {\bibinfo  {journal} {Nature Physics}\ }\textbf {\bibinfo {volume} {8}},\ \bibinfo {pages} {858} (\bibinfo {year} {2012})}\BibitemShut {NoStop}%
\bibitem [{\citenamefont {Fermi}(1949)}]{Fermi}%
  \BibitemOpen
  \bibfield  {author} {\bibinfo {author} {\bibfnamefont {E.}~\bibnamefont {Fermi}},\ }\href {\doibase 10.1103/PhysRev.75.1169} {\bibfield  {journal} {\bibinfo  {journal} {Phys. Rev.}\ }\textbf {\bibinfo {volume} {75}},\ \bibinfo {pages} {1169} (\bibinfo {year} {1949})}\BibitemShut {NoStop}%
\bibitem [{\citenamefont {Ho}\ and\ \citenamefont {Kulsrud}(1987)}]{Ho1987}%
  \BibitemOpen
  \bibfield  {author} {\bibinfo {author} {\bibfnamefont {D.~D.-M.}\ \bibnamefont {Ho}}\ and\ \bibinfo {author} {\bibfnamefont {R.~M.}\ \bibnamefont {Kulsrud}},\ }\href {\doibase 10.1063/1.866395} {\bibfield  {journal} {\bibinfo  {journal} {The Physics of Fluids}\ }\textbf {\bibinfo {volume} {30}},\ \bibinfo {pages} {442–461} (\bibinfo {year} {1987})}\BibitemShut {NoStop}%
\bibitem [{\citenamefont {Rudakov}\ and\ \citenamefont {Tsytovich}(1971)}]{Rudakov1971}%
  \BibitemOpen
  \bibfield  {author} {\bibinfo {author} {\bibfnamefont {L.~I.}\ \bibnamefont {Rudakov}}\ and\ \bibinfo {author} {\bibfnamefont {V.~N.}\ \bibnamefont {Tsytovich}},\ }\href {\doibase 10.1088/0032-1028/13/3/004} {\bibfield  {journal} {\bibinfo  {journal} {Plasma Physics}\ }\textbf {\bibinfo {volume} {13}},\ \bibinfo {pages} {213} (\bibinfo {year} {1971})}\BibitemShut {NoStop}%
\bibitem [{\citenamefont {Chaudhary}\ \emph {et~al.}(2018)\citenamefont {Chaudhary}, \citenamefont {Imam}, \citenamefont {Rizvi},\ and\ \citenamefont {Ali}}]{Chaudhary2018}%
  \BibitemOpen
  \bibfield  {author} {\bibinfo {author} {\bibfnamefont {K.}~\bibnamefont {Chaudhary}}, \bibinfo {author} {\bibfnamefont {A.~M.}\ \bibnamefont {Imam}}, \bibinfo {author} {\bibfnamefont {S.~Z.~H.}\ \bibnamefont {Rizvi}}, \ and\ \bibinfo {author} {\bibfnamefont {J.}~\bibnamefont {Ali}},\ }\enquote {\bibinfo {title} {Plasma kinetic theory},}\ in\ \href {\doibase 10.5772/intechopen.70843} {\emph {\bibinfo {booktitle} {Kinetic Theory}}}\ (\bibinfo  {publisher} {InTech},\ \bibinfo {year} {2018})\BibitemShut {NoStop}%
\bibitem [{\citenamefont {Paul~Drake}(2018)}]{PaulDrake2018}%
  \BibitemOpen
  \bibfield  {author} {\bibinfo {author} {\bibfnamefont {R.}~\bibnamefont {Paul~Drake}},\ }\href {\doibase 10.1088/1741-4326/aaf0e3} {\bibfield  {journal} {\bibinfo  {journal} {Nuclear Fusion}\ }\textbf {\bibinfo {volume} {59}},\ \bibinfo {pages} {035001} (\bibinfo {year} {2018})}\BibitemShut {NoStop}%
\bibitem [{\citenamefont {Vinogradova}\ and\ \citenamefont {Silkina}(2024)}]{Silkina}%
  \BibitemOpen
  \bibfield  {author} {\bibinfo {author} {\bibfnamefont {O.~I.}\ \bibnamefont {Vinogradova}}\ and\ \bibinfo {author} {\bibfnamefont {E.~F.}\ \bibnamefont {Silkina}},\ }\href {https://arxiv.org/abs/2312.02624} {\enquote {\bibinfo {title} {Conductivity of concentrated salt solutions},}\ } (\bibinfo {year} {2024}),\ \Eprint {http://arxiv.org/abs/2312.02624} {arXiv:2312.02624 [physics.chem-ph]} \BibitemShut {NoStop}%
\bibitem [{\citenamefont {Bonneau}\ \emph {et~al.}(2024)\citenamefont {Bonneau}, \citenamefont {Avni}, \citenamefont {Andelman},\ and\ \citenamefont {Orland}}]{Andelman}%
  \BibitemOpen
  \bibfield  {author} {\bibinfo {author} {\bibfnamefont {H.}~\bibnamefont {Bonneau}}, \bibinfo {author} {\bibfnamefont {Y.}~\bibnamefont {Avni}}, \bibinfo {author} {\bibfnamefont {D.}~\bibnamefont {Andelman}}, \ and\ \bibinfo {author} {\bibfnamefont {H.}~\bibnamefont {Orland}},\ }\href {https://arxiv.org/abs/2408.17427} {\enquote {\bibinfo {title} {Frequency-dependent conductivity of concentrated electrolytes: A stochastic density functional theory},}\ } (\bibinfo {year} {2024}),\ \Eprint {http://arxiv.org/abs/2408.17427} {arXiv:2408.17427 [cond-mat.soft]} \BibitemShut {NoStop}%
\bibitem [{\citenamefont {Tajima}\ and\ \citenamefont {Dawson}(1979)}]{Eubank}%
  \BibitemOpen
  \bibfield  {author} {\bibinfo {author} {\bibfnamefont {T.}~\bibnamefont {Tajima}}\ and\ \bibinfo {author} {\bibfnamefont {J.~M.}\ \bibnamefont {Dawson}},\ }\href {\doibase 10.1103/PhysRevLett.43.267} {\bibfield  {journal} {\bibinfo  {journal} {Phys. Rev. Lett.}\ }\textbf {\bibinfo {volume} {43}},\ \bibinfo {pages} {267} (\bibinfo {year} {1979})}\BibitemShut {NoStop}%
\bibitem [{\citenamefont {Cui}\ \emph {et~al.}(2017)\citenamefont {Cui}, \citenamefont {Milkus},\ and\ \citenamefont {Zaccone}}]{Cui2017}%
  \BibitemOpen
  \bibfield  {author} {\bibinfo {author} {\bibfnamefont {B.}~\bibnamefont {Cui}}, \bibinfo {author} {\bibfnamefont {R.}~\bibnamefont {Milkus}}, \ and\ \bibinfo {author} {\bibfnamefont {A.}~\bibnamefont {Zaccone}},\ }\href {\doibase 10.1103/PhysRevE.95.022603} {\bibfield  {journal} {\bibinfo  {journal} {Phys. Rev. E}\ }\textbf {\bibinfo {volume} {95}},\ \bibinfo {pages} {022603} (\bibinfo {year} {2017})}\BibitemShut {NoStop}%
\bibitem [{\citenamefont {Willis}\ \emph {et~al.}(2013)\citenamefont {Willis}, \citenamefont {Hagness},\ and\ \citenamefont {Knezevic}}]{Knezevich}%
  \BibitemOpen
  \bibfield  {author} {\bibinfo {author} {\bibfnamefont {K.~J.}\ \bibnamefont {Willis}}, \bibinfo {author} {\bibfnamefont {S.~C.}\ \bibnamefont {Hagness}}, \ and\ \bibinfo {author} {\bibfnamefont {I.}~\bibnamefont {Knezevic}},\ }\href {\doibase 10.1063/1.4798658} {\bibfield  {journal} {\bibinfo  {journal} {Applied Physics Letters}\ }\textbf {\bibinfo {volume} {102}},\ \bibinfo {pages} {122113} (\bibinfo {year} {2013})},\ \Eprint {http://arxiv.org/abs/https://pubs.aip.org/aip/apl/article-pdf/doi/10.1063/1.4798658/13220140/122113\_1\_online.pdf} {https://pubs.aip.org/aip/apl/article-pdf/doi/10.1063/1.4798658/13220140/122113\_1\_online.pdf} \BibitemShut {NoStop}%
\bibitem [{\citenamefont {Zaccone}(2023{\natexlab{a}})}]{viscosity}%
  \BibitemOpen
  \bibfield  {author} {\bibinfo {author} {\bibfnamefont {A.}~\bibnamefont {Zaccone}},\ }\href {\doibase 10.1103/PhysRevE.108.044101} {\bibfield  {journal} {\bibinfo  {journal} {Phys. Rev. E}\ }\textbf {\bibinfo {volume} {108}},\ \bibinfo {pages} {044101} (\bibinfo {year} {2023}{\natexlab{a}})}\BibitemShut {NoStop}%
\bibitem [{\citenamefont {Zaccone}(2023{\natexlab{b}})}]{zaccone2023theory}%
  \BibitemOpen
  \bibfield  {author} {\bibinfo {author} {\bibfnamefont {A.}~\bibnamefont {Zaccone}},\ }\href@noop {} {\emph {\bibinfo {title} {Theory of Disordered Solids: From Atomistic Dynamics to Mechanical, Vibrational, and Thermal Properties}}},\ Vol.\ \bibinfo {volume} {1015}\ (\bibinfo  {publisher} {Springer Nature},\ \bibinfo {address} {Cham},\ \bibinfo {year} {2023})\BibitemShut {NoStop}%
\bibitem [{\citenamefont {Kubo}(1966)}]{Kubo1966}%
  \BibitemOpen
  \bibfield  {author} {\bibinfo {author} {\bibfnamefont {R.}~\bibnamefont {Kubo}},\ }\href {\doibase 10.1088/0034-4885/29/1/306} {\bibfield  {journal} {\bibinfo  {journal} {Reports on Progress in Physics}\ }\textbf {\bibinfo {volume} {29}},\ \bibinfo {pages} {255} (\bibinfo {year} {1966})}\BibitemShut {NoStop}%
\end{thebibliography}%

\end{document}